\documentclass[twocolumn,showpacs,preprintnumbers,amsmath,amssymb]{revtex4-1}

\usepackage{graphicx}
\usepackage{dcolumn}
\usepackage{bm}
\usepackage{multirow}
\usepackage{float}
\usepackage{pdfpages}

\makeatletter
\providecommand \@ifxundefined [1]{%
 \@ifx{#1\undefined}
}%
\providecommand \@ifnum [1]{%
 \ifnum #1\expandafter \@firstoftwo
 \else \expandafter \@secondoftwo
 \fi
}%
\providecommand \@ifx [1]{%
 \ifx #1\expandafter \@firstoftwo
 \else \expandafter \@secondoftwo
 \fi
}%
\providecommand \natexlab [1]{#1}%
\providecommand \enquote  [1]{``#1''}%
\providecommand \bibnamefont  [1]{#1}%
\providecommand \bibfnamefont [1]{#1}%
\providecommand \citenamefont [1]{#1}%
\providecommand \href@noop [0]{\@secondoftwo}%
\providecommand \href [0]{\begingroup \@sanitize@url \@href}%
\providecommand \@href[1]{\@@startlink{#1}\@@href}%
\providecommand \@@href[1]{\endgroup#1\@@endlink}%
\providecommand \@sanitize@url [0]{\catcode `\\12\catcode `\$12\catcode
  `\&12\catcode `\#12\catcode `\^12\catcode `\_12\catcode `\%12\relax}%
\providecommand \@@startlink[1]{}%
\providecommand \@@endlink[0]{}%
\providecommand \url  [0]{\begingroup\@sanitize@url \@url }%
\providecommand \@url [1]{\endgroup\@href {#1}{\urlprefix }}%
\providecommand \urlprefix  [0]{URL }%
\providecommand \selectlanguage [0]{\@gobble}%
\providecommand \bibinfo  [0]{\@secondoftwo}%
\providecommand \bibfield  [0]{\@secondoftwo}%
\providecommand \BibitemOpen [0]{}%
\providecommand \BibitemShut  [1]{\csname bibitem#1\endcsname}%
\let\auto@bib@innerbib\@empty

\begin{document}

\preprint{}

\title{Electron-phonon coupling and thermal transport in the thermoelectric compound $\bm{\mathrm{Mo_3Sb_{7-x}Te_x}}$ 
}

\author{Dipanshu Bansal,$^1$ Chen W. Li,$^1$  Ayman H. Said,$^2$ Douglas L. Abernathy,$^3$ Jiaqiang Yan,$^{1,4}$ and Olivier Delaire$^1$}
\email{bansald@ornl.gov}
\email{delaireoa@ornl.gov}
\affiliation{
$^1$Materials Science and Technology Division, Oak Ridge National Laboratory, Oak Ridge, Tennessee 37831, USA \\
$^2$Advanced Photon Source, Argonne National Laboratory, Argonne, Illinois 60439, USA \\
$^3$Quantum Condensed Matter Division, Oak Ridge National Laboratory, Oak Ridge, Tennessee 37831, USA \\
$^4$Department of Materials Science and Engineering, University of Tennessee, Knoxville, Tennessee 37996, USA
}


\textbf{Notice: This manuscript has been authored by
UT-Battelle, LLC under Contract No. DE-AC05-
00OR22725 with the U.S. Department of Energy. The
United States Government retains and the publisher, by
accepting the article for publication, acknowledges that
the United States Government retains a non-exclusive,
paid-up, irrevocable, world-wide license to publish or
reproduce the published form of this manuscript, or
allow others to do so, for United States Government
purposes. The Department of Energy will provide
public access to these results of federally sponsored research
in accordance with the DOE Public Access Plan
(http://energy.gov/downloads/doe-public-access-plan).}

\begin{abstract}
Phonon properties of $\mathrm{Mo_3Sb_{7-x}Te_x}$ ($x=0,1.5, 1.7$), a potential high-temperature thermoelectric material, have been studied with inelastic neutron and x-ray scattering, and with first-principles simulations. The substitution of Te for Sb leads to pronounced changes in the electronic structure, local bonding, phonon density of states (DOS), dispersions, and phonon lifetimes. Alloying with tellurium shifts the Fermi level upward, near the top of the valence band, resulting in a strong suppression of electron-phonon screening, and a large overall stiffening of interatomic force-constants. The suppression in electron-phonon coupling concomitantly increases group velocities and suppresses phonon scattering rates, surpassing the effects of alloy-disorder scattering, and resulting in a surprising increased lattice thermal conductivity in the alloy. We also identify that the local bonding environment changes non-uniformly around different atoms, leading to variable perturbation strengths for different optical phonon branches.  The respective roles of changes in phonon group velocities and phonon lifetimes on the lattice thermal conductivity are quantified. Our results highlight the importance of the electron-phonon coupling on phonon mean-free-paths in this compound, and also estimates the contributions from boundary scattering, umklapp scattering, and point-defect scattering.

\end{abstract}

\maketitle


\section{Introduction}
Thermoelectric materials convert temperature gradients into electrical potentials, and are of widespread interest for their applications in  solid-state power generation or refrigeration \cite{Snyder2008, Minnich2009, Biswas2012, Keppens1998, Delaire2011, Nielsen2013}. Good thermoelectric conversion efficiency is dependent on achieving  large values of Seebeck coefficient, $S$, high electrical conductivity, $\sigma$, and at the same time a low thermal conductivity, $\kappa$. The total thermal conductivity is the sum of an electrical contribution from charge carriers, $\kappa_{\rm el}$, and a lattice component $\kappa_{\rm L}$ due to phonon propagation. In order to maximize the thermoelectric figure-of-merit, $ZT = S^2 \sigma T / (\kappa_{\rm el} + \kappa_{\rm L})$, much effort is focused toward suppressing phonon propagation to minimize  $\kappa_{\rm L}$ \cite{Snyder2008, Poudel2008, Minnich2009, Biswas2012, Keppens1998, Koza2008, Christensen2008, Ma2013, Voneshen2013, Rhyee2009}. 

$\mathrm{Mo_3Sb_{7}}$ is an over-doped $p$-type metallic compound and in order to improve its thermoelectric performance, one needs to dope extra electrons to reduce its carrier concentration \cite{Shi_2011,Parker_2011}. Adding electrons shifts the Fermi level toward the top of the valence band and increases the power factor (defined by product of square of Seebeck coefficient and electrical conductivity, $S^2\sigma$). Prior studies investigating electrical transport in $\mathrm{Mo_3Sb_{7-x}Te_x}$ show a definite  increase in power factor up to x = 1.8 \cite{Shi_2011,Parker_2011,Xu_2009,Candolfi_2009_1,Candolfi_2009_2,Candolfi_2009_3,Candolfi_2010}. Such alloying would normally be expected to reduce $\kappa_{\rm L}$ through impurity scattering \cite{Klemens,Tritt}. However, the introduction of Te in $\mathrm{Mo_3Sb_{7}}$ leads instead to a surprising increase in $\kappa_{\rm L}$ \cite{Shi_2011,Candolfi_2009_1,Candolfi_2009_2,Candolfi_2010}. This effect was attributed to a suppression in electron-phonon coupling as the hole concentration is reduced with Te doping \cite{Candolfi_2011, Shi_2011}. 

$\mathrm{Mo_3Sb_{7}}$ has also attracted strong interest because of its unusual magnetic and superconducting properties \cite{Bukowski_2002,Koyama_2008,Koyama_2009,Tran_2008,Yan_2013,Candolfi_2007,Wiendlocha_2008,Okabe_2009,Wiendlocha_2014}. $\mathrm{Mo_3Sb_{7}}$ undergoes a magnetic and structural phase transition at $T\simeq 53\,$K, from a  high-temperature cubic phase  ($Im\bar{3}m$) to a low-temperature tetragonal phase ($I4/mmm$), corresponding to a contraction of Mo-Mo nearest-neighbor bonds along the $c-$axis, and the formation of singlet dimers.  However, the distortion from the cubic structure is slight, with the $a/c$ ratio reaching 1.002 at 11\,K according to single-crystal x-ray diffraction \cite{Yan_2015}. Doping with Te  strongly suppresses the structural phase transition temperature, and no phase transition is observed for Te concentrations exceeding 0.36\% \cite{Yan_2015}. 

Prior inelastic neutron scattering (INS) investigations at low temperature ($T\le 300K$) have revealed a possible deviation from the quasi-harmonic (QH) approximation in the case of $\mathrm{Mo_3Sb_{7}}$, as opposed to a normal QH behavior in $\mathrm{Mo_3Sb_{5.4}Te_{1.6}}$ \cite{Candolfi_2011}. However, that study was limited to powder measurements of the phonon DOS at few temperatures. While Shi et al.~\cite{Shi_2011} investigated the thermal transport properties at high temperatures using transport and calorimetry measurements, momentum-resolved spectroscopic information about the phonon dispersions is still lacking. Here, we provide a detailed analysis of the effect of electron-phonon coupling by combining measurements of the phonon dispersions and density of states with x-ray and neutron scattering, and density functional theory (DFT) simulations including alloying effects explicitly. 

In the present study, we performed a comprehensive set of phonon measurements, complemented with first-principles simulations, which enable us to quantitatively assess the effect of Te alloying on the lattice dynamics and phonon transport in $\mathrm{Mo_3Sb_{7-x}Te_x}$. We investigated the phonon DOS of powders with INS in $\mathrm{Mo_3Sb_{7-x}Te_x}$ ($x=0,1.7$) for 12 temperatures over a wide range ($6 \leqslant T \leqslant 750\,$K). In addition, we performed inelastic x-ray scattering (IXS) measurements of low-energy phonon dispersions on single crystals in $\mathrm{Mo_3Sb_{7-x}Te_x}$ ($x=0,1.5$) at $T=100,300\,$K. We also performed density functional calculations of phonons for both the parent compound and Te-doped supercells representing the alloy, and gained insights into changes in bonding.  We combined our experiments and simulations to model the contributions of different phonon scattering mechanisms to $\kappa_{\rm L}$, which we compare with the reported thermal conductivity data. Our results clearly show that the electron-phonon coupling strongly affects the lattice dynamics and thermal conductivity in $\mathrm{Mo_3Sb_{7}}$, and is suppressed with Te alloying. Our analysis also points to an increase in phonon scattering owing to the coupled magnetic-structural fluctuations around $53\,$K in the undoped compound, although the magnetism does not appreciably alter the overall force-constants and phonon energies.

\section{Sample Preparation}
Polycrystalline $\mathrm{Mo_3Sb_{7-x}Te_x}$ ($x=0,1.7$) samples were synthesized by a solid-state reaction method. A homogeneous mixture (15\,grams) of Mo (Alfa, 99.999\%), Sb (Alfa, 99.9999\%), and Te (Alfa, 99.9999\%) powder was loaded into a quartz tube and sealed under approximately 1/3 atmosphere of argon gas. The Mo powder was first reduced in flowing Ar balanced with 4\% H2 for overnight at 1273\,K. The sealed ampoules were heated to 993\,K in 12 hours and kept at this temperature for three weeks. Room temperature x-ray powder diffraction measurements confirmed that both compositions had no impurity and were single-phase (cubic $Im\bar{3}m$). Furthermore, relative lattice parameter contraction in doped composition in comparison to undoped composition confirmed that the Te content is close to the nominal one. The superconducting transition temperature of $T_c = 2.30$\,K was confirmed for the parent compound by measuring magnetic susceptibility using a Quantum Design Magnetic Property Measurement System. No sign of superconductivity at 1.80\,K was observed in the powder sample with nominal composition of $\mathrm{Mo_3Sb_{5.3}Te_{1.7}}$.

Millimeter sized $\mathrm{Mo_3Sb_{7-x}Te_x}$ ($x=0,1.5$) single crystals were grown with a self-flux method technique \cite{Yan_2013,Yan_2015}. The real composition of as-grown crystals was determined by elemental analysis performed with a Hitachi TM-3000 tabletop electron microscope equipped with a Bruker Quantax 70 energy dispersive x-ray system. $\mathrm{Mo_3Sb_7}$ is the only line compound in Mo-Sb binary system, and our elemental analysis indicated no deviation from stoichiometry. The real composition of the Te-doped crystals was determined to be $\mathrm{Mo_3Sb_{5.48}Te_{1.52}}$. $\mathrm{Mo_3Sb_7}$ single crystals used in this study have been well characterized, as previously reported \cite{Yan_2013}. Magnetic measurements confirmed that $\mathrm{Mo_3Sb_{5.48}Te_{1.52}}$ single crystals do not superconduct above 1.8\,K.

\section{Inelastic Scattering Experiments}

\subsection{Neutron Scattering}

Inelastic neutron scattering measurements on $\mathrm{Mo_3Sb_{7-x}Te_x}$ ($x=0, 1.7$) powders were carried out using the ARCS time-of-flight chopper spectrometer at the Spallation Neutron Source (SNS) at Oak Ridge National Laboratory \cite{ARCS}. The powders (mass $\sim10$\,g) were encased in thin-walled aluminum cans.  Measurements at low temperature ($5 \le T \le 300$\,K) were performed with the sample container inside a closed-cycle helium refrigerator, with the sample chamber filled with a low pressure of helium.  High-temperature ($300 \le T \le 750$\,K) measurements used a low-background resistive furnace. All measurements used an oscillating radial collimator to minimize scattering from the sample environment. The empty aluminum can was measured in identical conditions at all temperatures. Two incident neutron energies $E_i=30$\,meV and 55\,meV were used at each temperature, to provide higher resolution data for low-energy optic modes, and to probe the full spectrum, respectively. The energy resolution (full width at half maximum) for incident energies $E_i=30$\,meV and 55\,meV are $\sim$1.3\,meV and $\sim$2.2\,meV at elastic line, respectively, which reduces to $\sim$0.4\,meV and $\sim$1.2\,meV at energy transfer of 30\,meV.

The data were normalized by the total incident flux, and corrected for detector efficiency. The transformation of data from instrument coordinates to the physical momentum transfer, $Q$,  and energy transfer, $E$, was performed using the MANTID reduction software \cite{MANTID}. The scattering from the empty container was processed identically and subtracted from the sample measurements. The analysis of the phonon DOS was performed within the incoherent scattering approximation, and corrected for the effect of multiphonon scattering \cite{Fultz2011}. The data in the region of the elastic peak were removed, and a Debye-like quadratic energy dependence was used to extrapolate the DOS at low energy ($E < 2$ meV for $E_i = 30$\,meV and $E < 4$\,meV for $E_i=55\,$meV).

The phonon spectra derived from INS measurements are skewed by the relative neutron scattering strengths of the elements present in the sample (neutron-weighting, NW). The partial phonon DOS of each element is weighted differently, according to the ratio of that element's scattering cross-section ($\sigma$) and mass (m). The neutron weighting factors for Mo, Sb, and Te are ${\sigma_{Mo}}/{m_{Mo}} = 0.0595$, ${\sigma_{Sb}}/{m_{Sb}} = 0.0320$, and ${\sigma_{Te}}/{m_{Te}} = 0.0339$, respectively (in units of barns/amu). Consequently, the total experimentally measured phonon DOS, $g(E)$, for $\mathrm{Mo_3Sb_{7-x}Te_x}$ is over-weighted for the Mo vibrational contributions:
\begin{eqnarray}
g_{NW}(E) &=& [ 3 \times \frac{\sigma_{Mo}}{m_{Mo}} g_{Mo}(E) + (7-x) \times \frac{\sigma_{Sb}}{m_{Sb}} g_{Sb}(E)  \nonumber \\
&& + x \times \frac{\sigma_{Te}}{m_{Te}} g_{Te}(E) ]  /10 \, ,
\end{eqnarray}
\noindent where $g_{Mo}(E)$, $g_{Sb}(E)$, and $g_{Te}(E)$ are the partial densities of states of Mo, Sb, and Te. However, the resulting neutron-weighted phonon DOS can be directly compared against the simulated DOS to which the same neutron-weights are applied (see below). We note that in this study, we keep the overall and partial phonon DOS normalized to a unit integral. 

\subsection{X-ray Scattering}

\begin{figure}
\begin{center}
\hspace{0.30in}
\includegraphics[trim=0cm 0.0cm 0cm 0.0cm, clip=true, width=0.39\textwidth]{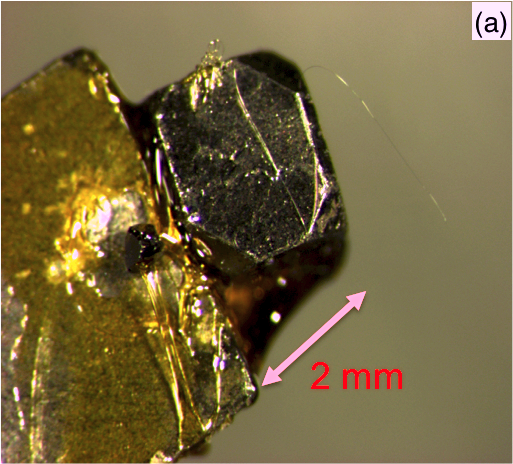}
\includegraphics[trim=0cm 0.0cm 0cm 0.0cm, clip=true, width=0.5\textwidth]{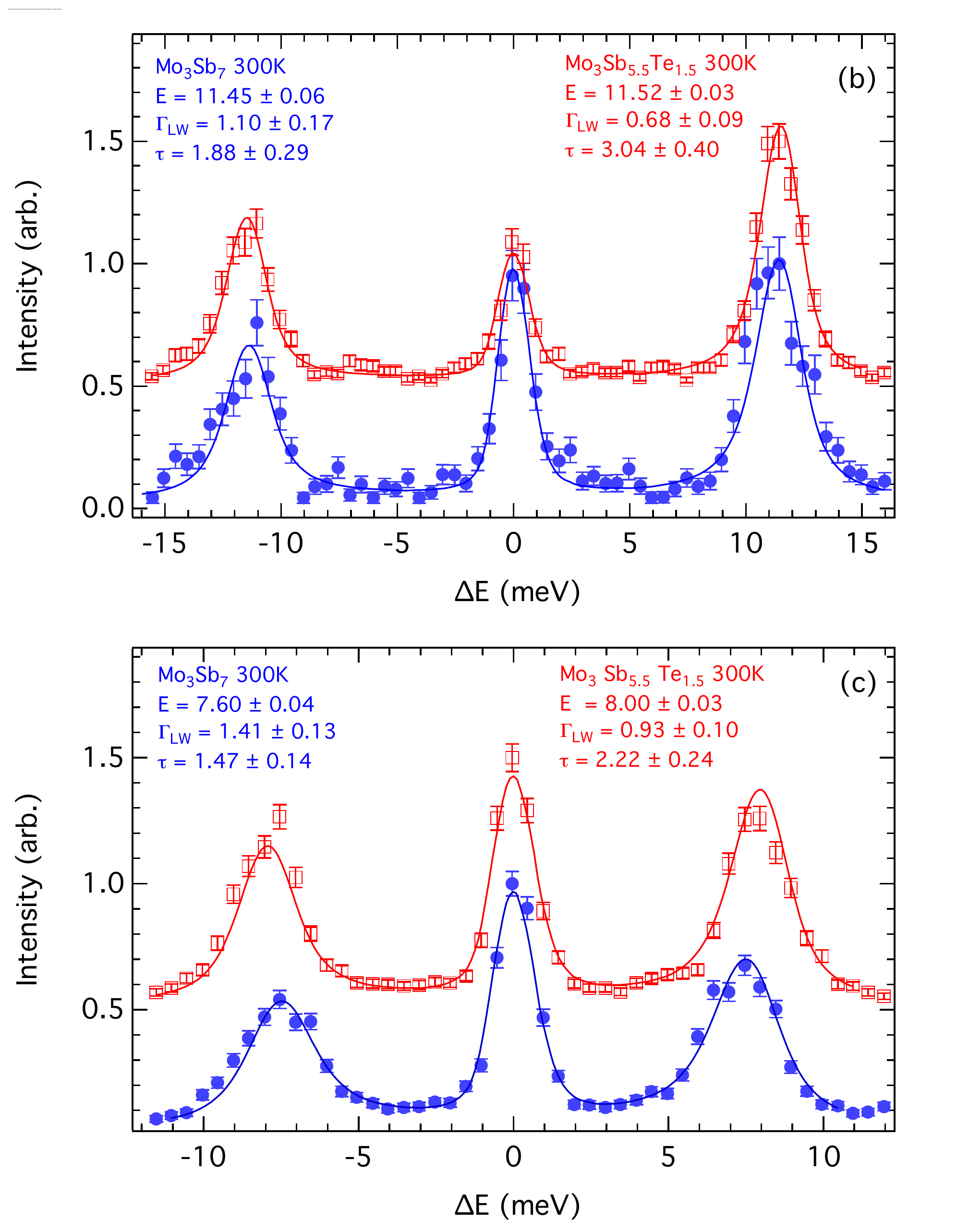}
\end{center}
\caption{\label{linewidth} (a) $\mathrm{Mo_3Sb_7}$ single crystal glued on beryllium post using varnish. The main facet, about 2\,mm across, corresponds to the (100) crystallographic plane. The very small crystal on the side of the beryllium mount was not used. (b) Constant-$\bm{Q}$ inelastic x-ray scattering spectra for longitudinal acoustic mode at $\bm Q = (0,6.6,0)$ for $\mathrm{Mo_3Sb_7}$ (filled blue circles) and $\mathrm{Mo_3Sb_{5.5}Te_{1.5}}$ (empty red squares) at 300\,K. The lines are fits using a damped harmonic oscillator profile convoluted with the instrument resolution (see text). The values of $E$ (in meV), $\Gamma_{LW}$ (in meV), and $\tau$ (in picosecond) listed in inset are the fitted phonon energy, phonon linewidth (corrected for resolution), and phonon lifetime, respectively. Negative $\Delta E$ corresponds to phonon annihilation, and positive $\Delta E$ to phonon creation. (c) Same as (b) for $\bm Q = (0,6.4,0)$.}
\end{figure}

Phonon dispersion curves of $\mathrm{Mo_3Sb_{7-x}Te_x}$ ($x=0, 1.5$) were extracted from the high resolution inelastic x-ray scattering (IXS) measurements on small single-crystals, using beamline 30-ID-C (HERIX \cite{HERIX1,HERIX2}) at the Advanced Photon Source (APS). The highly monochromatic x-ray beam had an energy  $E \simeq 23.7\,$keV ($\lambda = 0.5226\,$\AA), with an energy resolution $\Delta E \sim 1.0\,$meV (full width at half maximum), and was focused on the sample with a beam cross-section $\sim 35 \times 15\,\mu$m (horizontal$\times$vertical).  The convoluted energy resolution of monochromatic x-ray beam and analyzer crystal was $\Delta E \sim 1.5\,$meV. The single crystals of $\mathrm{Mo_3Sb_{7}}$ and $\mathrm{Mo_3Sb_{5.5}Te_{1.5}}$ were approximately $\sim 2\,$mm across with well defined facets. The measurements were performed in reflection geometry off $(100)$ facets, with the crystals glued to Beryllium posts using varnish, and mounted in a closed-cycle helium refrigerator. A crystal used for IXS is shown in figure~\ref{linewidth}-a. Typical counting times were in the range of 40 to 120\,s at each point in energy scans at constant $\bm{Q}$. 

The orientation matrix was defined using (6,6,0) and (0,6,0) Bragg peaks, and checked with other peaks. The longitudinal and transverse acoustic dispersions along high symmetry directions were measured along $(0,6+\xi,0)$, $(0,6,\xi)$, $(6+\xi,6+\xi,0)$, $(6-\xi,6+\xi,0)$, $(\xi,6,-\xi)$, and $(\xi,6-\xi,-\xi)$. The spectra were fitted with a damped-harmonic-oscillator scattering function \cite{Lovesey1984}, {subsequently convoluted with the measured instrumental resolution function, $R * S(E)$}:
\begin{align}
S(E) = A\,\frac{\{\frac{1}{2} \pm \frac{1}{2} + n(|E|)\} \times \frac{1}{2}\Gamma_{LW}}{(E^2 - E_0^2)^2 + (\frac{1}{2}\Gamma_{LW})^2} + B\,,
\end{align}
where $B$ is the constant background, $n(E)$ is the temperature-dependent Bose-Einstein distribution at phonon energy transfer $E$, $A$ is the amplitude, $\Gamma_{LW}$ is full phonon linewidth at half maximum, and $E_0$ is the bare phonon energy in the absence of damping (the $+$ sign is for $E>0$ and $-$ sign for $E<0$). Typical IXS spectra and the corresponding fits are shown in figure~\ref{linewidth}.  

\section{Density Functional Theory Simulations}

First-principles simulations were performed in the framework of density functional theory (DFT) as implemented in the Vienna Ab initio Simulation Package (VASP 5.3) \cite{VASP_link, Kresse1993, Kresse1996b, Kresse1996}.  A $4\times4\times4$ Monkhorst Pack electronic \emph{k}-point mesh was used for all of our simulations, with a plane-wave cut-off energy of 500\,eV. The projector-augmented-wave potentials explicitly included 14 valence electrons for Mo ($4s^24p^65s^14d^5$), 5 for Sb ($5s^25p^3$), and 6 for Te ($5s^25p^4$). All  our calculations used the generalized gradient approximation with PBE parametrization \cite{refPBE}. 

During the relaxation of the structure, the lattice parameters were kept fixed at the experimental value for the cubic phase at room temperature \cite{Candolfi_2008_1,Candolfi_2009_1}, and atomic positions were optimized until forces on all atoms were smaller than 1\,meV\,\AA$^{-1}$. The calculations of phonon dispersions were performed in the harmonic approximation, using the finite displacement approach as implemented in Phonopy \cite{Phonopy_link, Phonopy}, with the atomic forces in the distorted supercells obtained with VASP. In the case of $\mathrm{Mo_3Sb_{7}}$, the phonon calculations used a $2\times 2\times 2$ supercell of the primitive cell, containing 160 atoms. 

\begin{figure*}
\begin{center}
\includegraphics[trim=0cm 11.0cm 0cm 0.0cm, clip=true, width=0.9\textwidth]{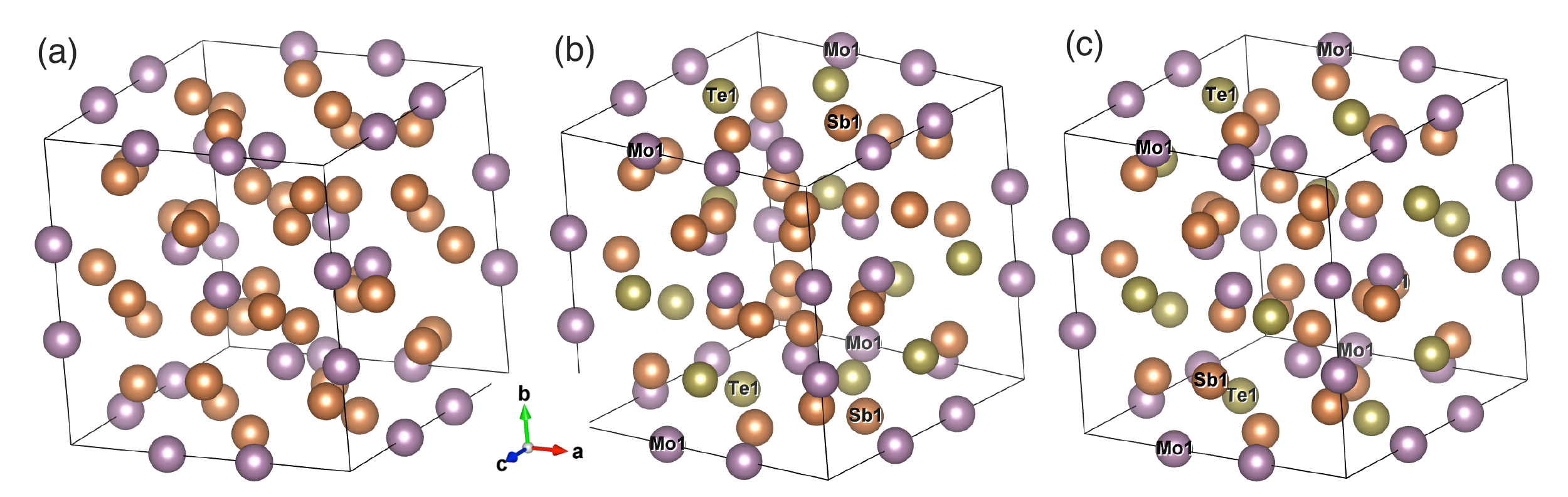}
\end{center}
\caption{\label{unitcell} Unitcell of a) $\mathrm{Mo_3Sb_7}$, b) $\mathrm{Mo_3Sb_{5.5}Te_{1.5}}$ Case1, and c) $\mathrm{Mo_3Sb_{5.5}Te_{1.5}}$ Case2 (Mo atoms: purple color, Sb atoms: orange color, and Te atoms: light green color). Owing to strong site preference of Te atom in $\mathrm{Mo_3Sb_{7-x}Te_{x}}$, only Sb(1) atoms at 12d site were replaced with Te atoms. Case1 and Case2 represent two possible Te doped configurations. }
\end{figure*}

To describe the random alloy with $x=1.5$, we built two cubic supercells (40 atoms), in which six Sb atoms were replaced with six Te atom to represent Mo:Sb:Te :: 3:5.5:1.5 composition as $\mathrm{Mo_{12}Sb_{22}Te_{6}}$.  The cubic unit cells of $\mathrm{Mo_3Sb_{7}}$ and the two cells representing the alloy $\mathrm{Mo_3Sb_{5.5}Te_{1.5}}$ are shown in figure~\ref{unitcell}. In the body-centered-cubic structure (BCC) of $\mathrm{Mo_3Sb_{7}}$, the Mo, Sb(1), and Sb(2) atoms occupy crystallographic sites 12e, 12d, and 16f. Prior studies of $\mathrm{Mo_3Sb_{7-x}Te_{x}}$ have shown, using neutron diffraction and ab initio computations of energetics, that Te substitution occurs preferentially at the 12d site as opposed to 16f \cite{Candolfi_2008_1}. Hence, in building the 40-atom supercells describing the alloy, we have replaced 6 Sb(1) atoms at site 12d with 6 Te atoms. There are multiple  possible choices for placing six Te atoms on the Sb(1) sites. To limit the computational requirements, we chose  two configurations (case 1 and case 2), illustrated in figure~\ref{unitcell}. To compute phonons for the alloy, $2\times 2\times 2$ supercells of the two alloy cells in each case (the two phonon supercells contained 320 atoms). To construct the force-constant matrix using the finite displacement approach, we computed 5, 26 and 36 independent atomic displacements for the undoped cell, case 1, and case 2, respectively. The atomic displacement amplitude was one hundredth of the lattice constant. 

\begin{figure}
\begin{center}
\includegraphics[trim=5cm 0.0cm 35cm 0.0cm, clip=true, width=0.5\textwidth]{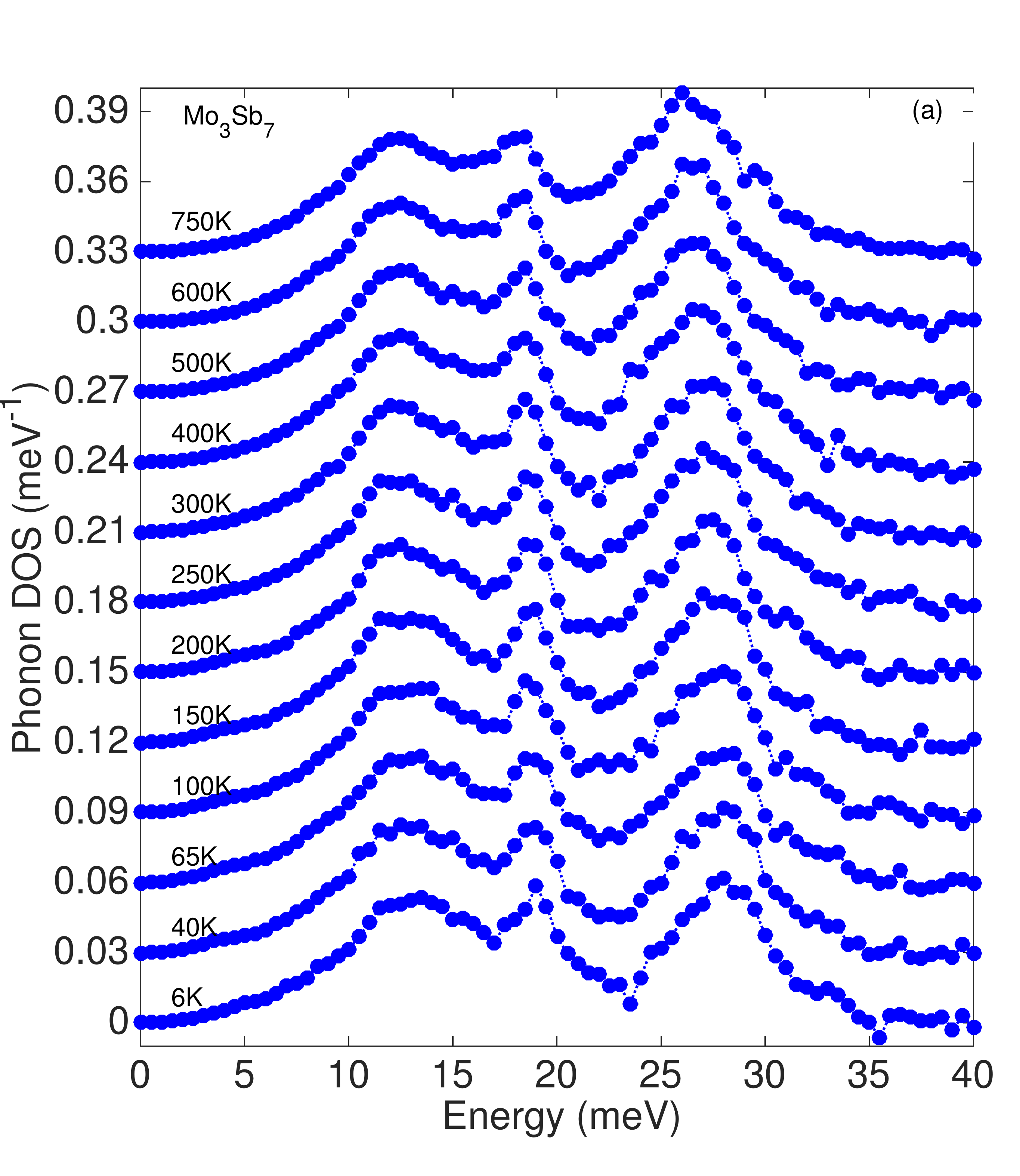}
\includegraphics[trim=34.8cm 0.0cm 5.1cm 0.0cm, clip=true, width=0.5\textwidth]{figure3-eps-converted-to.pdf}
\end{center}
\caption{\label{DOS_Mo3Sb7} Neutron-weighted phonon DOS of $\mathrm{Mo_3Sb_7}$ measured with inelastic neutron scattering at different temperatures for incident neutron energies of a) 55\,meV and b) 30\,meV. }
\end{figure}

\begin{figure}
\begin{center}
\includegraphics[trim=5cm 0.0cm 35cm 0.0cm, clip=true, width=0.5\textwidth]{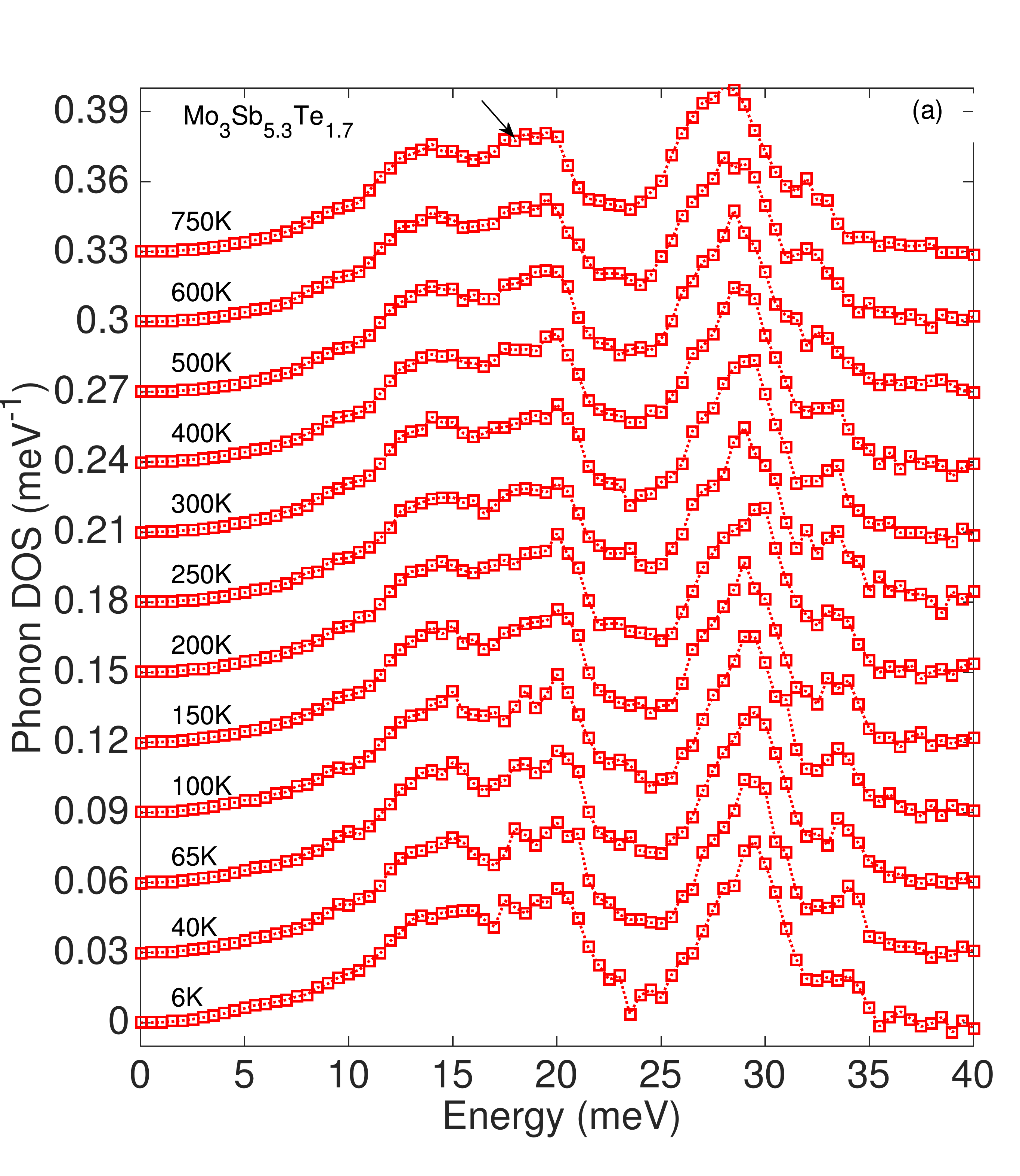}
\includegraphics[trim=34.8cm 0.0cm 5.1cm 0.0cm, clip=true, width=0.5\textwidth]{figure4-eps-converted-to.pdf}
\end{center}
\caption{\label{DOS_doped} Neutron-weighted phonon DOS of $\mathrm{Mo_3Sb_{5.3}Te_{1.7}}$ measured with inelastic neutron scattering at different temperatures for incident neutron energies of a) 55\,meV and b) 30\,meV. The peak denoted by arrow is described in text.}
\end{figure}

\section{Results and Discussion}

\subsection{Phonon Density of States}

\begin{figure*}
\begin{center}
\includegraphics[trim=3.5cm 13.5cm 3.5cm 0.0cm, clip=true, width=1.0\textwidth]{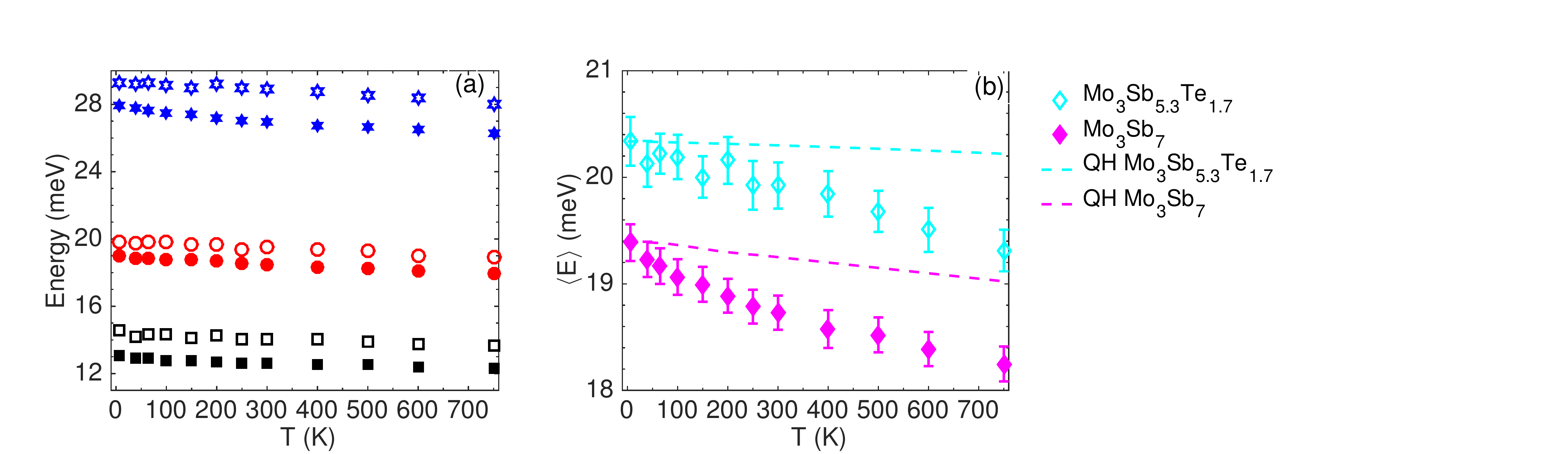}
\includegraphics[trim=3.5cm 13.5cm 3.5cm 1.0cm, clip=true, width=1.0\textwidth]{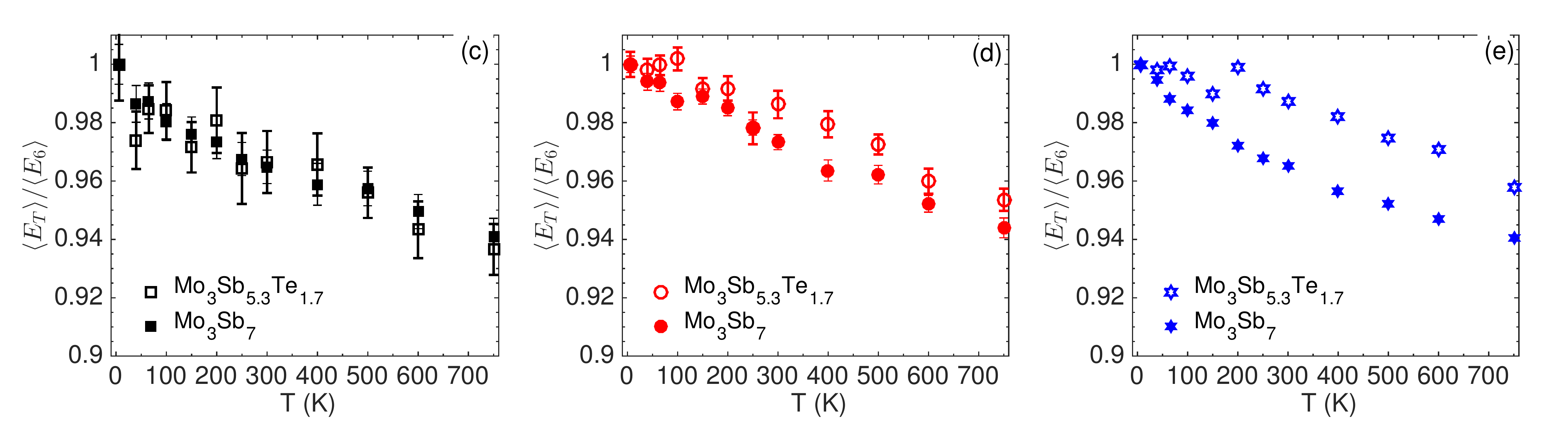}
\includegraphics[trim=3.5cm 13.5cm 3.5cm 1.0cm, clip=true, width=1.0\textwidth]{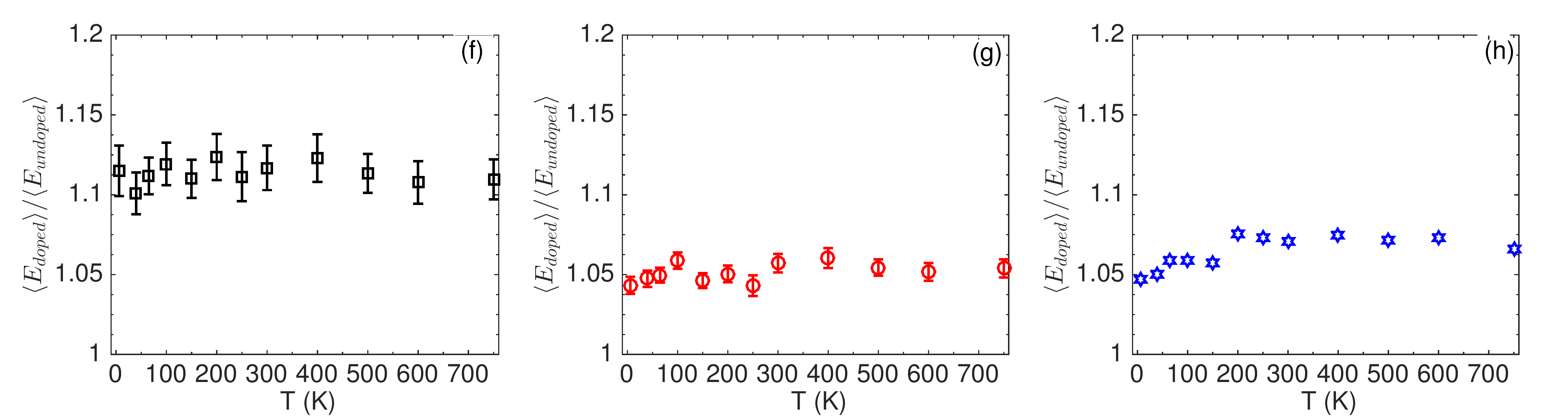}
\end{center}
\caption{\label{phonon_softening} a) Softening of phonon peaks at $\sim$14, 19, and 28\,meV $\mathrm{Mo_3Sb_7}$ (filled symbol) and $\mathrm{Mo_3Sb_{5.3}Te_{1.7}}$ (empty symbol) measured from inelastic neutron scattering at different temperatures for incident neutron energies of 55 meV, b) Comparison of phonon mean energy $\langle E\rangle$ of $\mathrm{Mo_3Sb_7}$ and $\mathrm{Mo_3Sb_{5.3}Te_{1.7}}$. The dashed curves show the expected temperature dependence in the quasiharmonic model (QH), using the experimental data of thermal Gruneisen parameters and volumetric expansion. Moreover, the relative change in phonon peak energy c) $\sim$14\,meV, d) $\sim$19\,meV, and e) $\sim$28\,meV with temperature ${\langle E_T\rangle}/{\langle E_{6\rm K}\rangle}$, and doping ${\langle E_{\rm doped}\rangle}/{\langle E_{\rm undoped}\rangle}$ of phonon peaks at f) $\sim$14\,meV, g) $\sim$19\,meV, and h) $\sim$28\,meV, respectively. In some panels, experimental error bars are comparable to the size of the symbols, and omitted.}
\end{figure*}

\begin{figure}
\begin{center}
\includegraphics[trim=4cm 9.5cm 4cm 9.0cm, clip=true, width=0.45\textwidth]{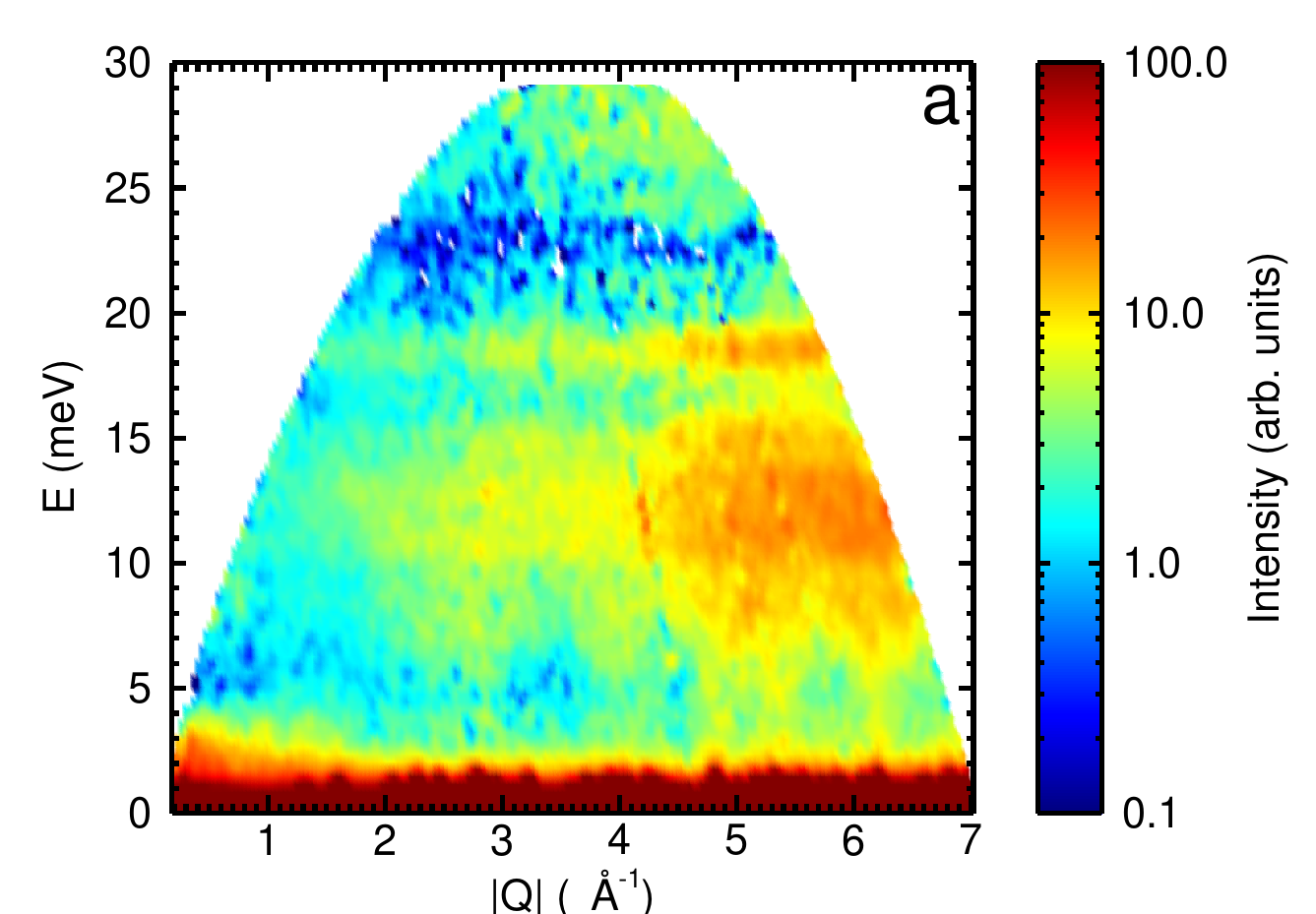}
\includegraphics[trim=4cm 9.5cm 4cm 9.0cm, clip=true, width=0.45\textwidth]{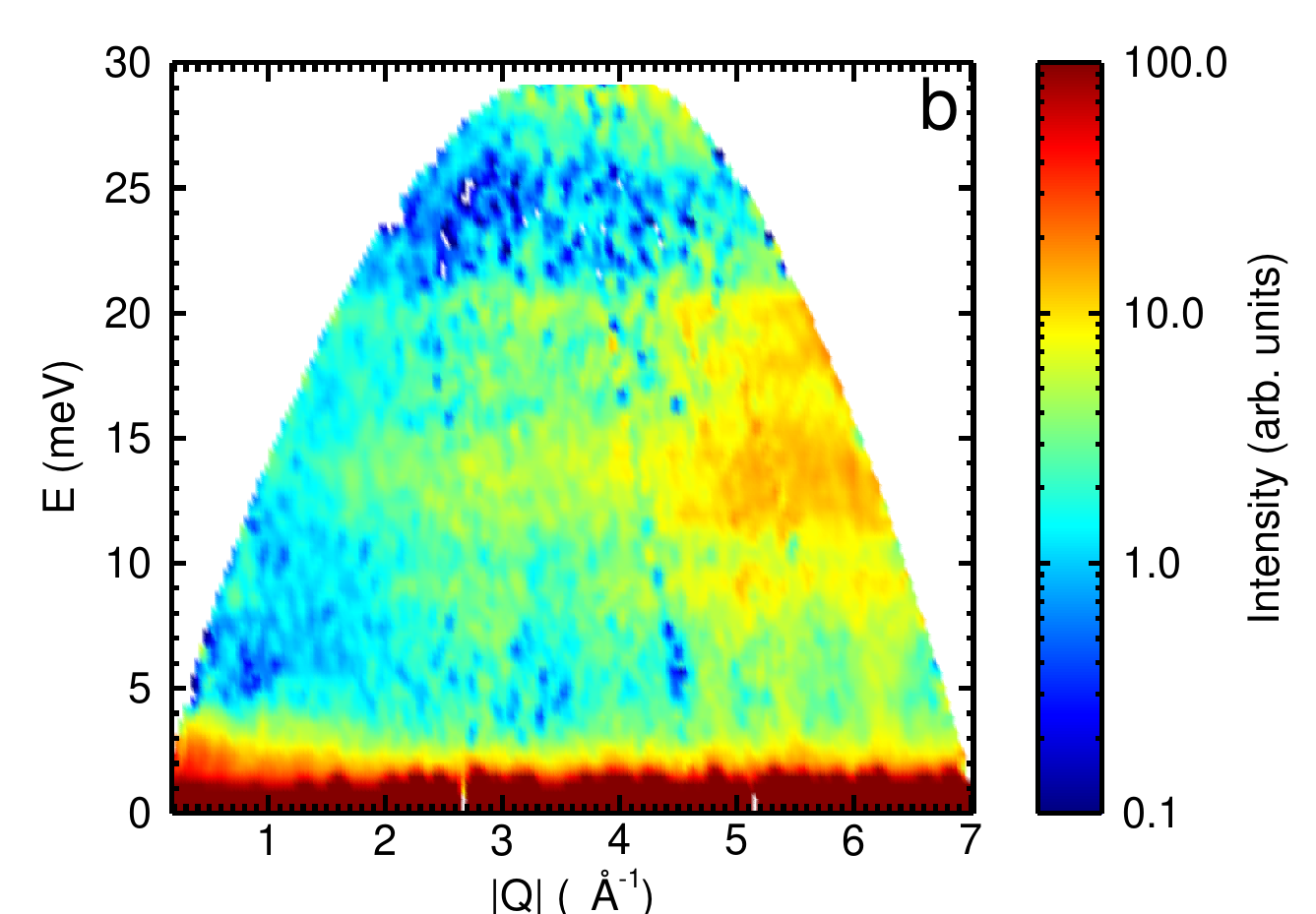}
\vspace{-0.1in}
\end{center}
\caption{\label{SQE} Experimental scattering intensity function $S({\bm Q},E)$ of a) $\mathrm{Mo_3Sb_7}$ and b) $\mathrm{Mo_3Sb_{5.3}Te_{1.7}}$ measured with inelastic neutron scattering at 6\,K for incident neutron energies of 30\,meV. At low momentum transfers $|Q|$, we could not detect any magnetic scattering intensity in $\mathrm{Mo_3Sb_7}$.}
\end{figure}

The neutron-weighted phonon DOS from INS are shown in figures~\ref{DOS_Mo3Sb7} and~\ref{DOS_doped} for $\mathrm{Mo_3Sb_7}$ and $\mathrm{Mo_3Sb_{5.3}Te_{1.7}}$, respectively, at the different temperatures measured and for the two incident energies used. A clear softening of phonons (shift to lower frequencies) is observed over the full phonon DOS when increasing temperature, but with quite different magnitudes in the doped and undoped compounds. To analyze this softening, we have compared the mean phonon energy $\langle E\rangle$ with the expected softening upon thermal expansion within the quasi-harmonic (QH) approximation. We use an average thermodynamic Gr\"{u}neisen parameter $\gamma = {3\alpha VK_T}/{C_v}$, where $\alpha$ is coefficient of linear thermal expansion, $V$ is the unit cell volume, $K_T$ the iso-thermal bulk modulus, and $C_v$ the heat capacity at constant volume (per unit cell). The iso-thermal bulk modulus was obtained from the phonon group velocities at 300\,K derived from our IXS measurements (table~\ref{elastic_modulus}), {while $C^{ph}_v$ was calculated from the phonon DOS measured at T=6\,K, in the harmonic approximation, and magnetic and electronic contributions to the specific heat were added at low temperatures \cite{Tran_2008}. We obtain $\gamma = 1.25$ and 0.70 for $\mathrm{Mo_3Sb_7}$ and $\mathrm{Mo_3Sb_{5.3}Te_{1.7}}$, respectively.} It can be observed from figure~\ref{phonon_softening}-b, that, while both materials deviate from the QH approximation for $T>300\,$K, this deviation is more pronounced in the undoped system at low temperatures $T\le 200K$. While $\mathrm{Mo_3Sb_{5.3}Te_{1.7}}$ follows the QH behavior fairly well at low $T$, $\mathrm{Mo_3Sb_7}$ displays an unusual temperature dependence and softens much faster, in agreement with the data reported by Candolfi \textit{et al.}~\cite{Candolfi_2011}. As we will show later, this pronounced softening in $\mathrm{Mo_3Sb_7}$ is due to the electron-phonon coupling, which strongly influences phonon energies, lifetimes, and group velocities. We point out that the INS data on powders do not show much change across the 53\,K phase transition in $\mathrm{Mo_3Sb_7}$. We could not detect any magnetic scattering intensity at low $|Q|$ (see figure~\ref{SQE}). The NW phonon DOS curves also show little change between 40\,K and 65\,K (figure~\ref{DOS_Mo3Sb7}), indicating a relatively weak effect of magnetic correlations on the overall lattice dynamics (force-constants) in this system. However, the magnetic fluctuations and the distortions from pairing of Mo dimers are likely to lead to increased phonon scattering rates (increased linewidths), which would not be easily detectable in the phonon DOS. Our analysis of thermal transport based on phonon group velocities reveal this effect more clearly (see below). 

\begin{figure*}
\begin{center}
\includegraphics[trim=6.0cm 1.5cm 6.0cm 0.0cm, clip=true, width=0.95\textwidth]{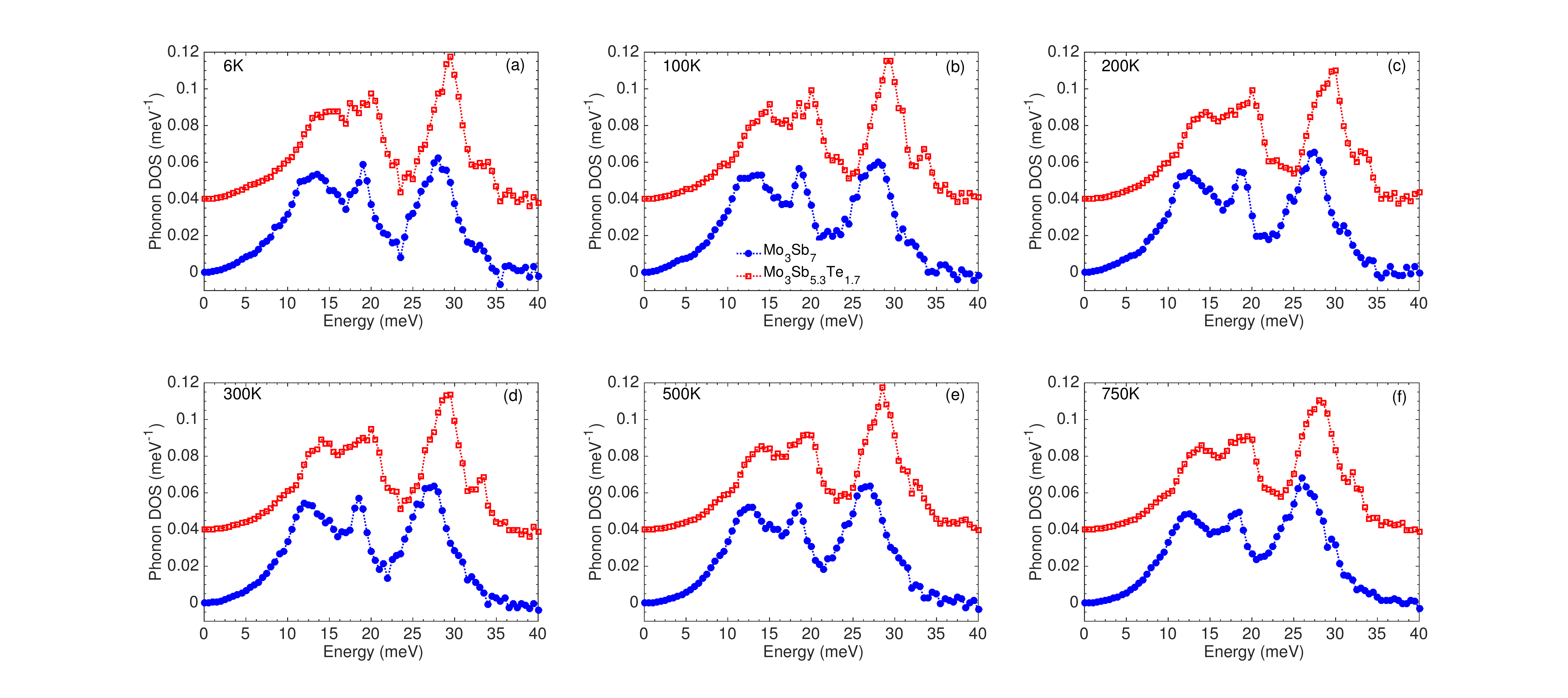}
\end{center}
\caption{\label{Comparison_DOS} Comparison of neutron-weighted phonon DOS of $\mathrm{Mo_3Sb_7}$ (filled symbol) $\mathrm{Mo_3Sb_{5.3}Te_{1.7}}$ (empty symbol) measured from inelastic neutron scattering at a) 6K, b) 100K, c) 200K, d) 300K, e) 500K, and f) 750K for incident neutron energy of 55\,meV. }
\end{figure*}

Figure~\ref{Comparison_DOS} illustrates the effect of Te alloying. A significant stiffening $\Delta E \sim$1.0\,meV is observed upon introduction of Te, and results from a combination of decrease in lattice parameter upon Te doping and of reduction in carrier concentration (from $\sim10^{22}$ $cm^{-3}$ in $\mathrm{Mo_3Sb_7}$ to $\sim10^{21}$ $cm^{-3}$ in $\mathrm{Mo_3Sb_{5.3}Te_{1.7}}$ \cite{Shi_2011}). {Importantly, the contraction in lattice parameter with Te doping ($\Delta a / a \sim0.1\%$ \cite{Candolfi_2008_1}) only accounts for 0.35\% of the observed phonon DOS stiffening. Thus, additional mechanisms such as the change in bonding and suppression in electron-phonon coupling could explain the large observed stiffening.} The stiffening with Te doping occurs at all temperatures and remains fairly constant with temperature (see figure~\ref{phonon_softening}(f, g, h)). 

\begin{figure*}
\begin{center}
\includegraphics[trim=5.5cm 1.5cm 5.5cm 0.0cm, clip=true, width=1.0\textwidth]{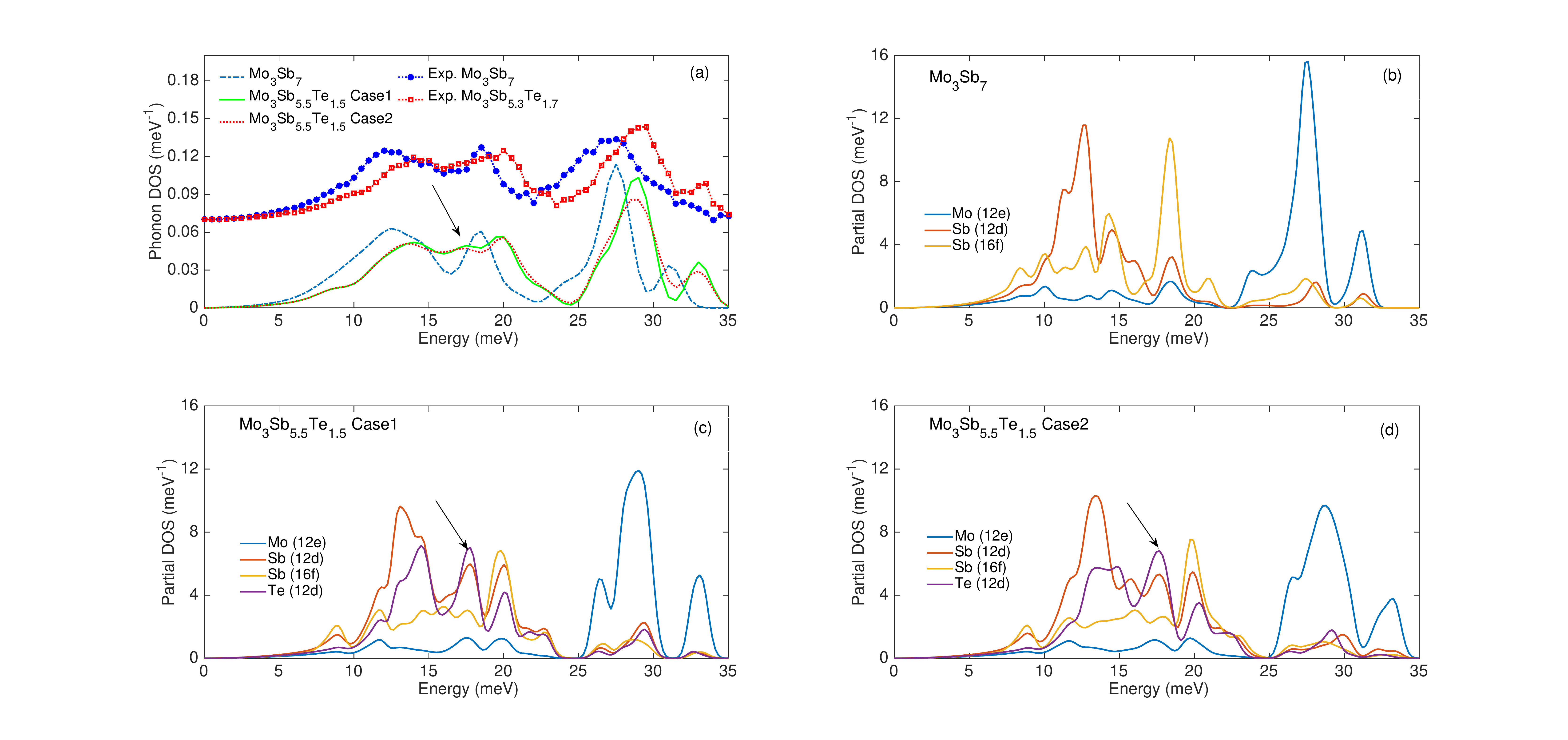}
\end{center}
\caption{\label{DOS} a) DFT calculated neutron weighted phonon DOS convoluted with ARCS resolution function at 55\,meV compared with experimental phonon measured at 300K, and b, c, d) partial phonon DOS  of $\mathrm{Mo_3Sb_7}$, $\mathrm{Mo_3Sb_{5.5}Te_{1.5}}$ Case1, and $\mathrm{Mo_3Sb_{5.5}Te_{1.5}}$ Case2, respectively, computed from first-principles. The peak denoted by an arrow corresponds to the Sb (12d) site, and is most sensitive to Te doping, as discussed in the text. }
\end{figure*}

Interestingly, by examining figure~\ref{Comparison_DOS}, one can observe that the magnitude of phonon energy shifts upon Te doping is not uniform across the entire DOS. In particular, the phonon peak at $\sim$14 meV is most sensitive to doping ($\Delta E / E \sim 10$\%) while the peaks at $\sim$19\,meV and $\sim$28\,meV stiffen by about 5\% and 6\%, respectively. We identified the origin of this difference by determining which atomic sites are predominantly involved in these different frequency regions of the phonon DOS, based on our DFT simulations. As can be seen in figure~\ref{DOS}, in the simulated partial phonon DOS, the modes at $\sim$14, 19, and 28\,meV correspond predominantly to vibrations of Sb(12d), Sb(16f), and Mo(12e) atoms, respectively. Thus, the 14\,meV peak is most sensitive to the Te doping, owing to the preferred Te substitution on the (12d) site. This behavior also reveals a larger bonding change around the Te dopants (12d), rather than an average renormalization of force-constants. We discuss this point in more details below.

\subsection{Bonding}

\begin{figure*}
\begin{center}
\includegraphics[trim=4.5cm 0.0cm 4.5cm 0.0cm, clip=true, width=0.95\textwidth]{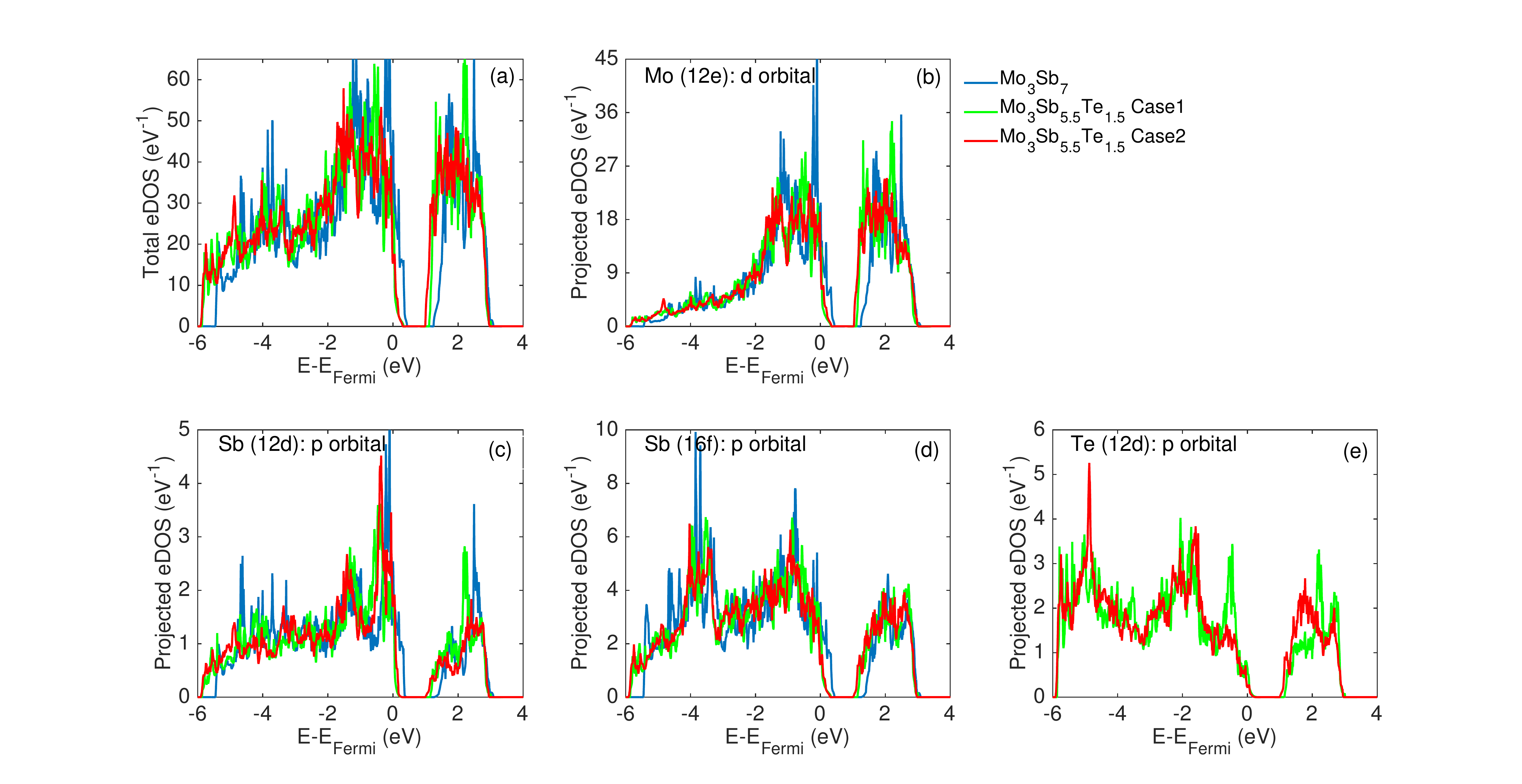}
\end{center}
\caption{\label{EDOS_all} Electronic density of states (eDOS) computed from first-principles for $\mathrm{Mo_3Sb_7}$ and $\mathrm{Mo_3Sb_{5.5}Te_{1.5}}$. (a) total eDOS.  (b, c, d, e) site-projected eDOS. }
\end{figure*}

To investigate the modifications in chemical bonding responsible for the observed changes in vibrational properties, we first examine the electronic density of states (eDOS) of $\mathrm{Mo_3Sb_7}$ and $\mathrm{Mo_3Sb_{5.5}Te_{1.5}}$ obtained with DFT, shown in figure~\ref{EDOS_all}. As can be seen in this figure, the two doped supercells (case 1 and case 2) have a very similar eDOS, indicating that the particular location of Te dopants among 12d sites does not impact the results strongly. In both of the doped cells, the Fermi level (origin of energy axis) is pushed up towards the top of the valence band, compared to $\mathrm{Mo_3Sb_7}$, with little distortion of the eDOS. This rigid band shift is rationalized as the filling of holes by the extra $d$ electrons of Te. Because of the slope in the eDOS, this shift of $E_{\rm F}$ decreases the number of electronic states at the Fermi level ($\sim42\%$), in agreement with previous observations by Candolfi \textit{et al.}~\cite{Candolfi_2008_1,Candolfi_2009_1} and Xu \textit{et al.}~\cite{Xu_2009}. { This decrease in eDOS at Fermi level originates from $d$-states of Mo(12e), and $p$-states of Sb(12d), Sb(16f), and Te(12d) atoms, with the projected eDOS showing suppressions at the Fermi level of about \begin{small}$\left\{\left(N_{Mo(12e)}^{undoped}(E_F) - N_{Mo(12e)}^{doped}(E_F)\right)/N_{Mo(12e)}^{undoped}(E_F)\right\}$$\sim$18\%, $\left\{\left(N_{Sb(12d)}^{undoped}(E_F) - N_{Sb(12d)}^{doped}(E_F)\right)/N_{Sb(12d)}^{undoped}(E_F)\right\}$$\sim$14\%, $\left\{\left(N_{Sb(16f)}^{undoped}(E_F) - N_{Sb(16f)}^{doped}(E_F)\right)/N_{Sb(16f)}^{undoped}(E_F)\right\}$$\sim$18\%\end{small}, and \begin{small}$\left\{\left(N_{Sb(12d)}^{undoped}(E_F) - N_{Te(12d)}^{doped}(E_F)\right)/N_{Sb(12d)}^{undoped}(E_F)\right\}$ $\sim$77\%\end{small}, respectively.} In order to evaluate these changes, we have averaged the eDOS of supercells in case 1 and case 2. 

In previous studies, some of us showed that sharp features in the eDOS near the Fermi level can result in a strong sensitivity of the electron-phonon coupling on either temperature or doping  \cite{Delaire_2008_PRL, Delaire_2008_PRB, Delaire_2009_PRB, Delaire_2011_PNAS}. A similar behavior is observed in the present case of $\mathrm{Mo_3Sb_7}$, with its strong sensitivity to band filling (Te doping). At low temperature, $\mathrm{Mo_3Sb_{5.3}Te_{1.7}}$ follows the QH model, while the significantly larger electronic density at the Fermi level in $\mathrm{Mo_3Sb_7}$ leads to a large deviations from QH behavior. 

The change in density at the Fermi level impacts the effectiveness of screening of forces on the ion cores. Our first-principles lattice dynamics simulations of $\mathrm{Mo_3Sb_7}$ and $\mathrm{Mo_3Sb_{5.5}Te_{1.5}}$ illustrate this effect very clearly. As we can observe from figure~\ref{DOS_Mo3Sb7},~\ref{DOS_doped}, and~\ref{DOS}, our simulations capture the lattice dynamics of the two compositions very well, including the overall shape of the spectrum, as well as the frequencies of phonon peaks, and also the relative stiffening with Te doping. In particular, we observe the pronounced shift of the Te-dominated peak at $\sim18$ meV in both simulations and measurements. A very good agreement is also achieved between our IXS measurements of  phonon dispersions and our DFT simulations, as will be discussed below. 

\begin{figure*}
\begin{center}
\includegraphics[trim=0cm 3.8cm 0cm 0.5cm, clip=true, width=0.9\textwidth]{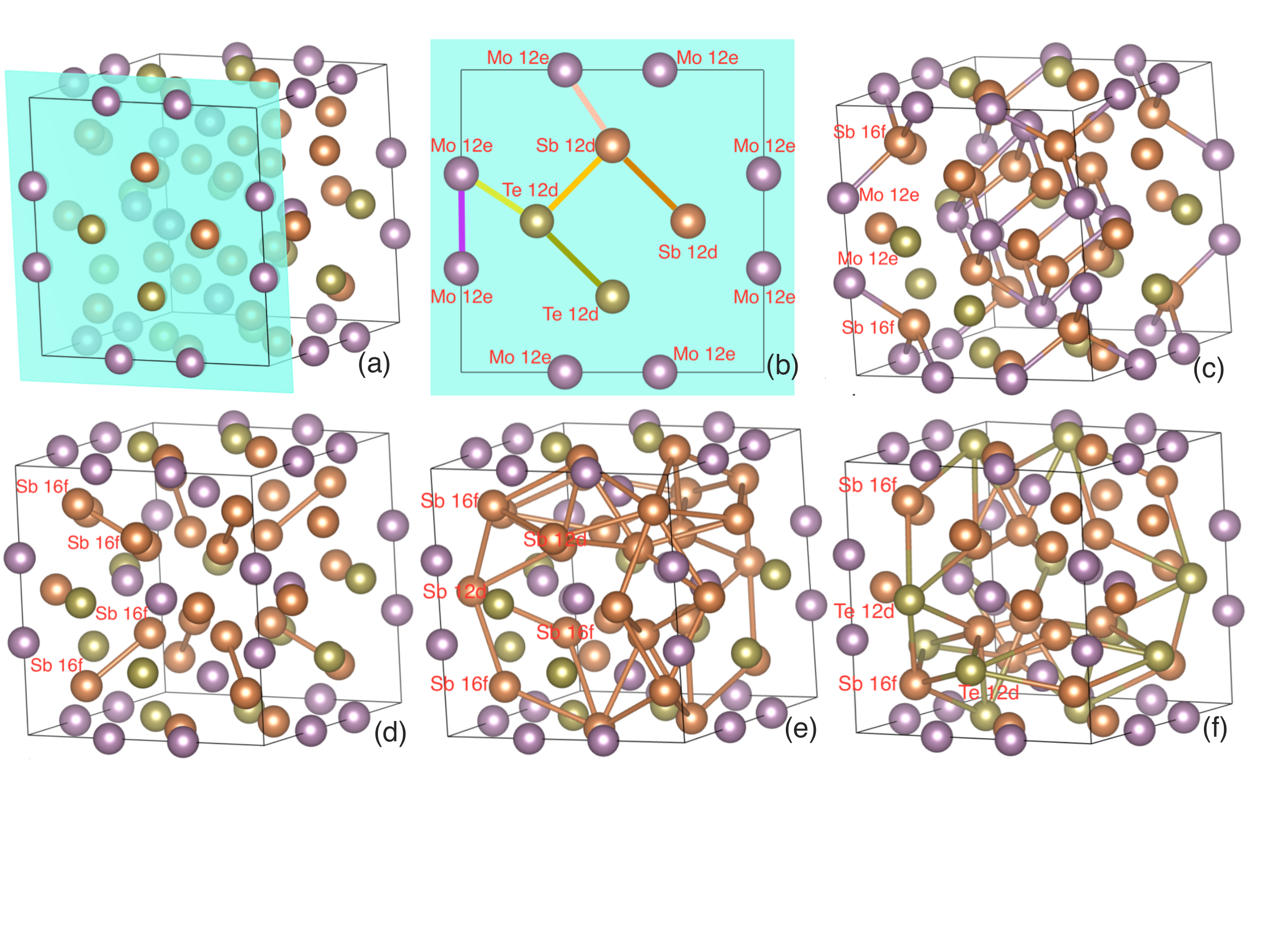}
\end{center}
\caption{\label{BvK_bonds} First-nearest-neighbor interatomic bonds between the different pairs of atoms considered in table~\ref{BvK_comparison} and in the text. a) and b) Mo(12e)-Mo(12e) , Mo(12e)-Sb(12d), Mo(12e)-Te(12d), Sb(12d)-Sb(12d) , Sb(12d)-Te(12d), and Te(12d)-Te(12d), c) Mo(12e)-Sb(16f), d) Sb(16f)-Sb(16f), e) Sb(12d)-Sb(16f), and f) Sb(16f)-Te(12d).}
\end{figure*}

\begin{figure}
\begin{center}
\includegraphics[trim=5.5cm 15.0cm 42.5cm 0.5cm, clip=true, width=0.4\textwidth]{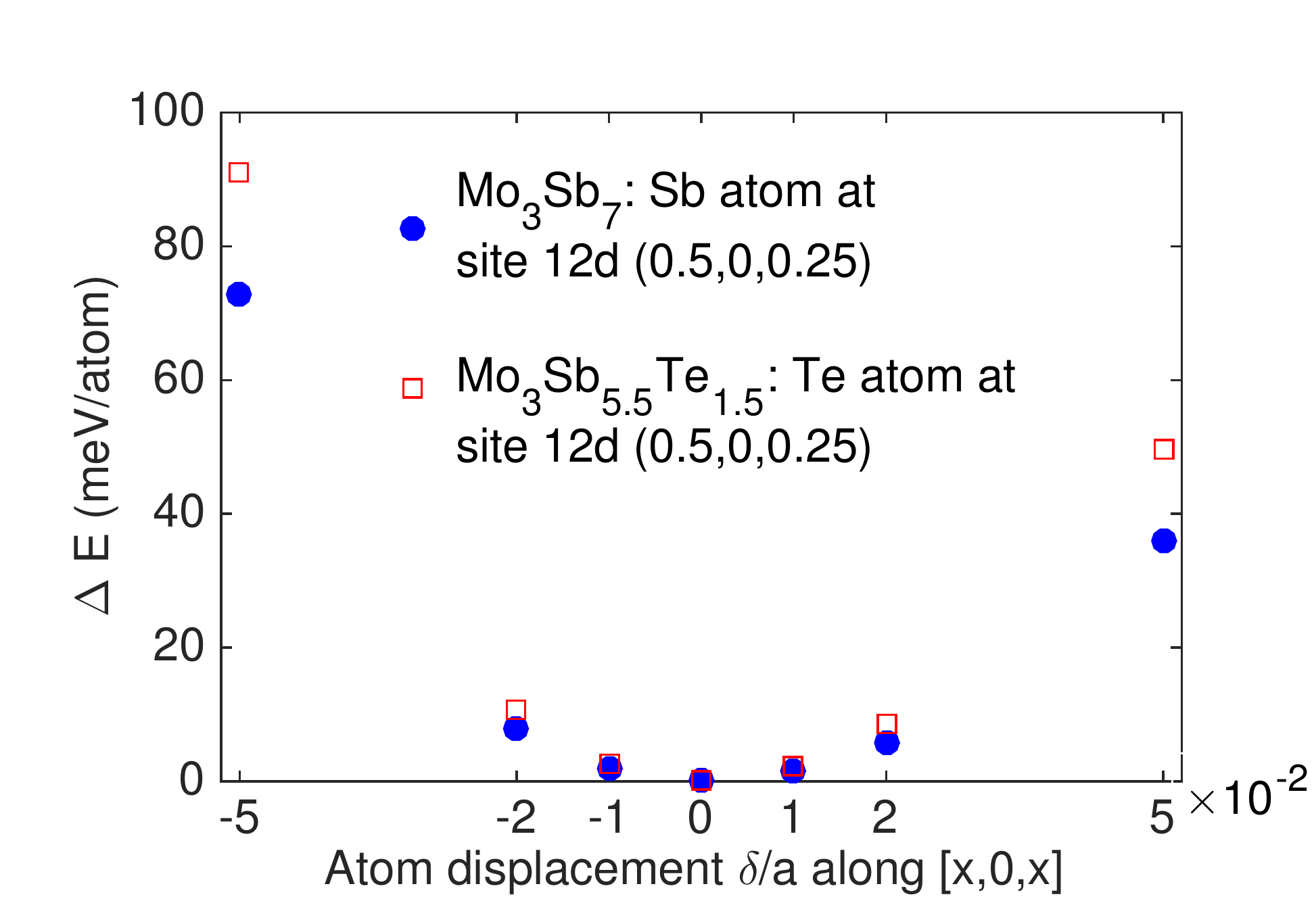}
\end{center}
\caption{\label{local_bonding} Comparison of change in DFT electronic energy of $\mathrm{Mo_3Sb_7}$ (filled symbol) and $\mathrm{Mo_3Sb_{5.5}Te_{1.5}}$ (empty symbol) while displacing Sb/Te atom at site 12d along [x,0,x], by a relative amount $\delta/a$. Here, $a$ is the lattice constant of $\mathrm{Mo_3Sb_7}$ and $\mathrm{Mo_3Sb_{5.5}Te_{1.5}}$ at 300\,K.}
\end{figure}

Given the reliability of our DFT simulations, we can rationalize the effects of Te doping observed experimentally by investigating the behavior of interatomic force-constants. Table~\ref{BvK_comparison} lists the Born-von Karman (BvK) force constants of two compositions for first-nearest-neighbor (1NN) bonds between different pairs of atoms (see illustration in figure~\ref{BvK_bonds}). The values reported in table~\ref{BvK_comparison} correspond to the $L_1$ norm of the $3\times3$ BvK force-constant matrix for each nearest pair of atoms. Since in the alloy supercell case 2, we replaced Sb with Te at 12d sites that are far apart, a direct comparison with 1NN force-constants is not possible. Thus, we limit ourselves to comparing case 1 with the undoped $\mathrm{Mo_3Sb_7}$ cell. As can be seen in table~\ref{BvK_comparison}, the increase in force-constants with Te doping is large, in particular for the bonds Mo(12e)-Mo(12e), Sb(12d)-Sb(12d), and Sb(16f)-Sb(16f). A slight decrease ($\sim6\%$) of the stiffness of the Mo(12e)-Sb(12d) 1NN bond is observed with Te doping. However, the Mo(12e)-Te(12d) bond is comparatively much stiffer, effectively stiffening the average Mo(12e)-Te/Sb(12d) bond in the alloy $\mathrm{Mo_3Sb_{5.5}Te_{1.5}}$, in comparison to Mo(12e)-Sb(12d) in the pure compound $\mathrm{Mo_3Sb_7}$. Accounting for the respective number of bonds in the alloy, we determine that the average Mo(12e)-Te/Sb(12d) bond stiffness increases by 108\%, compared to the undoped system. The large stiffening of this bond is reflected in the experimentally observed 10\% stiffening of the $\sim14$ meV phonon peak (figure~\ref{phonon_softening}), which is dominated by Sb/Te(12d) atomic motions, as seen in the partial phonon DOS (figure~\ref{DOS}). Similarly, the $\sim65\%$ stiffening of the bond stiffness for Mo(12e)-Mo(12e) and $\sim45\%$ stiffening for Sb(16f)-Sb(16f) lead to the observed $\sim6\%$ and $\sim5\%$ stiffening of phonon peaks at $\sim$28 and 19 meV, respectively. The change in local bonding upon Te substitution can also be seen by computing the stiffness of the local potential energy surface with DFT. Figure~\ref{local_bonding} shows the potential curves for small displacements of the Sb or Te atom from the 12d site, revealing a much stiffer bonding in the case of Te.

\begin{table*}
\caption{Comparison of first-nearest-neighbor Born-von Karman (BvK) force-constants for $\mathrm{Mo_3Sb_7}$ and $\mathrm{Mo_3Sb_{5.5}Te_{1.5}}$. The values are $L_1$ norm of $3\times3$ BvK force-constant matrices (in $\mathrm{eV/\mathring{A}^2}$) between nearest pairs of atoms.}
  \label{BvK_comparison}
\begin{center}
  \begin{tabular}{|c|c|c|c|}
  \hline
  BvK FC ($\mathrm{eV/\mathring{A}^2}$) & $\mathrm{Mo_3Sb_{5.5}Te_{1.5}}$ & $\mathrm{Mo_3Sb_7}$ & $\%$ increase \\
  \cline{2-4}
  \hline
Mo(12e)-Mo(12e)	&	6.16	&	3.76	&	63.70	\\
Mo(12e)-Sb(12d)	&	6.07	&	6.45	&	-5.78\\
Mo(12e)-Sb(16f)	&	13.59	&	11.63	&	16.79	\\
Mo(12e)-Te(12d)	&	7.53	&	$\cdots$	&	$\cdots$	\\
Sb(12d)-Sb(12d)	&	2.23	&	1.37	&	62.60	\\
Sb(12d)-Sb(16f)	&	1.91	&	1.63	&	17.48	\\
Sb(12d)-Te(12d)	&	2.85	&	$\cdots$	&	$\cdots$	\\
Sb(16f)-Sb(16f)	&	6.95	&	4.74	&	46.47	\\
Sb(16f)-Te(12d)	&	2.01	&	$\cdots$	&	$\cdots$	\\
Te(12d)-Te(12d)	&	3.45	&	$\cdots$	&	$\cdots$	\\
\hline
  \end{tabular}
  \end{center}
\end{table*}

Based on the BvK force-constants, one can rationalize the behavior of experimental and computed phonon spectra. Considering the force-constants of Sb(12d)-Sb(12d) (2.23 $\rm{eV/\mathring{A}^2}$) and Te(12d)-Te(12d) (3.45 $\rm{eV/\mathring{A}^2}$) in doped cells, we observe that the pair of Te(12d) atoms is much stiffer than the Sb(12d) pair. This separates the frequencies of Te(12d)-dominated vibrations from Sb(12d) vibrations, giving rise to two distinct phonon peaks at $\sim$14\,meV and $\sim$18\,meV. In contrast, the much smaller change in force-constants of Sb(12d)-Sb(16f) (1.91 $\rm{eV/\mathring{A}^2}$) and Te(12d)-Sb(16f) (2.01 $\rm{eV/\mathring{A}^2}$), and Sb(12d)-Mo(12e) (6.07 $\rm{eV/\mathring{A}^2}$) and Te(12d)-Mo(12e) (7.53 $\rm{eV/\mathring{A}^2}$) do not lead to any peak splitting (see figure~\ref{DOS}).

\subsection{Thermal Transport}

\begin{figure*}
\begin{center}
\includegraphics[trim=4cm 1.5cm 6cm 0.0cm, clip=true, width=0.9\textwidth]{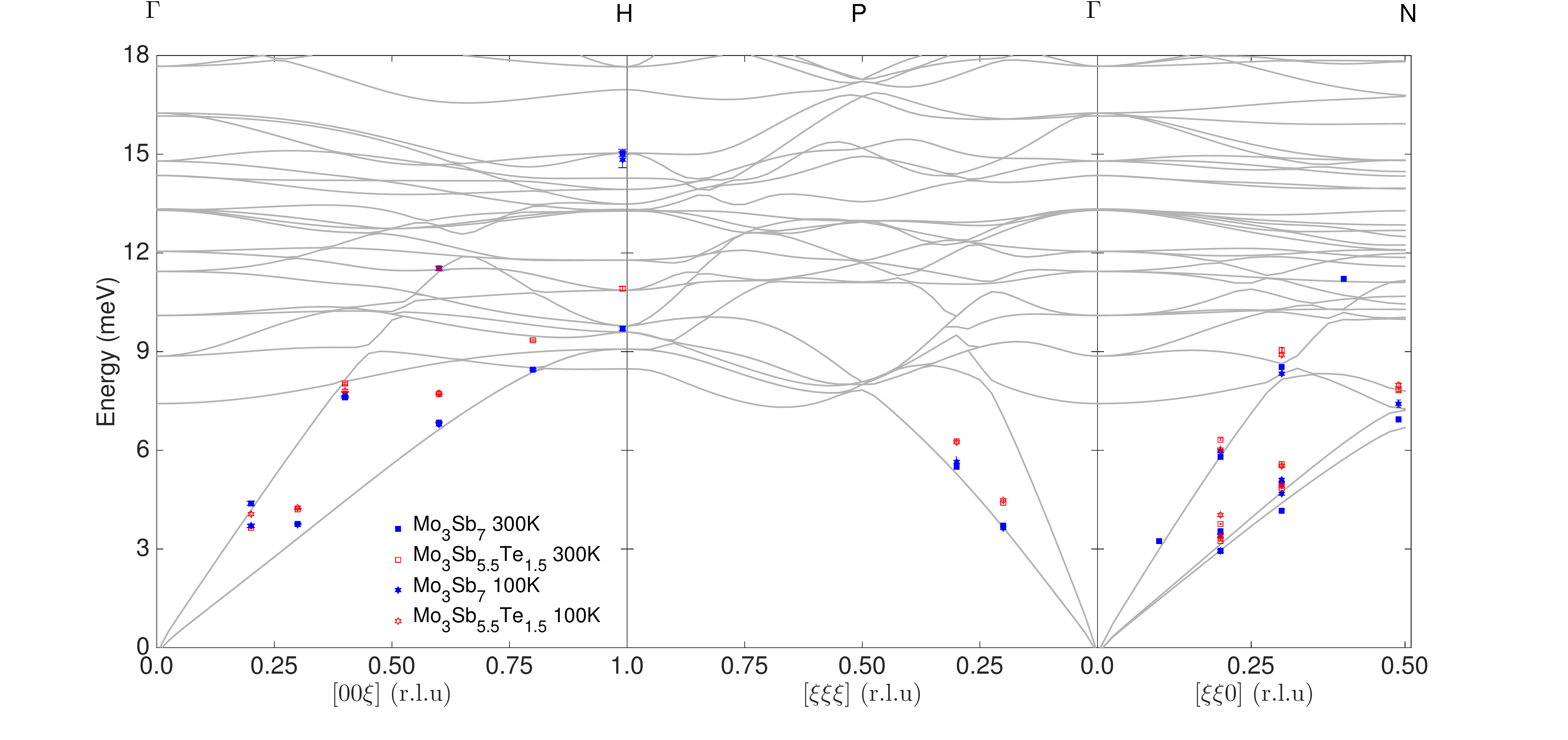}
\end{center}
\caption{\label{experimental_data} Phonon energies of $\mathrm{Mo_3Sb_7}$ (filled symbol) and $ \mathrm{Mo_3Sb_{5.5}Te_{1.5}}$ (empty symbol) obtained from experimental inelastic x-ray scattering measurements. The light-gray lines are DFT simulations of phonon dispersions for $\mathrm{Mo_3Sb_7}$ along high symmetry directions, which can be compared with the filled blue squares for IXS data ($\mathrm{Mo_3Sb_7}$ at 300\,K).}
\end{figure*}

In the remainder of the discussion, we focus on thermal transport properties. To quantify the impact of the change in bonding on the thermal conductivity, we consider the phonon dispersions for acoustic modes and low-energy optical branches. Figure~\ref{experimental_data} shows the phonon dispersions of $\mathrm{Mo_3Sb_7}$ computed with DFT (thin lines), together with experimental momentum-resolved phonon energies from IXS measurements (markers), for both $\mathrm{Mo_3Sb_7}$ and the Te-doped compound, at $T=100$\,K and 300\,K. The phonon energies were determined from the IXS spectra at specific wave vectors, using the fitting procedure described in introduction and illustrated in Fig.~\ref{linewidth}. We find a very good agreement between the IXS measurements for $\mathrm{Mo_3Sb_7}$ at 300\,K and the DFT simulations (which used experimental lattice parameters at 300\,K). Our DFT simulations for the phonon dispersions are in fair overall agreement with prior reports by Candolfi \textit{ et al.}~\cite{Candolfi_2011} and Wiendlocha \textit{ et al.}~\cite{Wiendlocha_2008,Wiendlocha_2014}, except for the strong underestimation of the energy of the transverse acoustic branch along $\mathrm{\Gamma-N}$ in these prior studies. This is likely the result of the smaller cells used to compute the phonons in these prior studies. The larger unit cells used in our computations were necessary to achieve the level of agreement with experiments observed here.

Considering the effect of Te alloying on phonon dispersions, we note that the amount of stiffening depends on the polarization of branches. The stiffening of transverse acoustic (TA) modes is pronounced along all directions measured ($\Gamma$--H, $\Gamma$--P, $\Gamma$--N). On the other hand, the LA mode stiffens noticeably with Te-doping along $\Gamma$--N, but there is little change of LA with doping, on average, along  $\Gamma$--H. Tables~\ref{gv_comparison} and~\ref{elastic_modulus} compare the phonon group velocities and elastic moduli derived from the phonon dispersions calculated with DFT and measured with IXS, as well as elastic moduli from resonant ultrasound measurements \cite{Lindsay}. The elastic moduli from phonon group velocity $v_g$ have been calculated  by solving $y = Hx$,
\begin{align}
y = \begin{Bmatrix} \rho v^2_{g:[00\xi]LA} \\
\rho v_{g:[00\xi]TA}^2 \\
2\rho v_{g-[\xi\xi0]LA}^2 \\
2\rho v_{g-[\xi\xi0]TA1}^2 \\
\rho v_{g-[\xi\xi0]TA2}^2 \\
3\rho v_{g-[\xi\xi\xi]TA}^2 \\
2\rho v_{g-[\xi\xi\xi]LA}^2 \\
\end{Bmatrix};
\hspace{0.05in}
H &= \begin{bmatrix}
1 & 0 & 0 \\
0 & 0 & 1 \\
1 & 1 & 2 \\
1 & -1 & 0 \\
0 & 0 & 1 \\
1 & -1 & 1 \\
1 & 2 & 4 \\
\end{bmatrix};\nonumber\\
\text{and}\hspace{0.05in}
x &= \begin{Bmatrix}
C_{11} \\
C_{12} \\
C_{44} \\
\end{Bmatrix}.
\end{align}
where $\rho = 8640\,kg/m^3$ is the density of $\mathrm{Mo_3Sb_7}$ \cite{MPDS}. The density of Te-doped composition is assumed to be same as undoped composition, as $\rm{Sb}$ and $\rm{Te}$ have similar atomic masses ($m_{Te}/m_{Sb}\sim1.05$), and lattice parameter are approximately equal (less than 0.1\% different at 300K). The bulk modulus, $(C_{11}+2C_{12})/3$, obtained from IXS dispersions is adiabatic $K_{ad}$, and can be converted to iso-thermal bulk modulus using ${K_T = K_{ad}}/(1+\alpha\gamma T)$. In all cases, we find that the change is practically negligible and within the experimental error bars, however (less than 0.075\%). The good agreement between experimental IXS data, resonant ultrasonic  measurements, and DFT simulations provides confidence in our methods and results. We note that our experimental ($K_T=112\pm 18$\,GPa) and computational results ($K_T$ = 134\,GPa) for the bulk modulus are much larger than the estimates ($K_T=$ 72--73.6\,GPa) reported by Candolfi \textit{et al.}~\cite{Candolfi_2011}

\begin{table*}
  \caption{Phonon group velocities ($v_g$ in m/s) of long-wavelength acoustic phonons for $\mathrm{Mo_3Sb_7}$ and $\mathrm{Mo_3Sb_{5.5}Te_{1.5}}$, calculated from DFT and measured with IXS. Phonon group velocities from IXS data has been estimated by fitting linear function to two $\bm{q}-$points along that particular direction with constraint -- phonon energy is zero at $\rm{\Gamma}$ point.}
  \label{gv_comparison}
  \begin{tabular}{|c|c|c|c|c|c|c|c|c|}
  \hline
\multirow{2}{*}{Direction} & \multirow{2}{*}{Mode} & \multicolumn{3}{c|}{$\mathrm{Mo_3Sb_7}$ $v_g$ (m/s)}  &  \multicolumn{4}{c|}{$\mathrm{Mo_3Sb_{5.5}Te_{1.5}}$ $v_g$ (m/s)}  \\
 \cline{3-9}
& &  DFT & Exp.~100K & Exp.~300K & Case1 DFT & Case2 DFT  & Exp.~100K & Exp.~300K \\
\hline
\multirow{2}{*}{$\mathrm{\Gamma-H}$}	&	TA	&	2578	&	2626$\pm$157	&	2650$\pm$152	&	3153	&	3180	&	2979$\pm$176	&	2975$\pm$156	\\
	&	LA	&	4857	&	4461$\pm$103	&	4408$\pm$386	&	4830	&	4380	&	4515$\pm$110	&	4636$\pm$254	\\
\hline
\multirow{3}{*}{$\mathrm{\Gamma-N}$}	&	TA1	&	2483	&	2538$\pm$107	&	2292$\pm$91	&	2228	&	2589	&	2704$\pm$59	&	2650$\pm$6	\\
	&	TA2	&	2609	&	2779$\pm$10	&	2736$\pm$132	&	3211	&	3193	&	3049$\pm$213	&	3046$\pm$24	\\
	&	LA	&	4929	&	4596$\pm$242	&	4737$\pm$335	&	5280	&	5334	&	4862$\pm$59	&	4972$\pm$177	\\
\hline
\multirow{2}{*}{$\mathrm{\Gamma-P}$}	&	TA	&	2501	&	2500$\pm$64	&	2455$\pm$26	&	2471	&	2821	&	2806$\pm$154	&	2816$\pm$111	\\
	&	LA	&	4923	&	$\cdots$	&	$\cdots$	&	5497	&	5409	&	$\cdots$	&	$\cdots$	\\
    \hline
  \end{tabular}
\end{table*}

\begin{table*}
  \caption{Comparison of elastic moduli $C_{11}, C_{12}, C_{44}$, and iso-thermal bulk modulus $K_T$ calculated from DFT and derived from slopes of IXS dispersions, for $\mathrm{Mo_3Sb_7}$ and $\mathrm{Mo_3Sb_{5.5}Te_{1.5}}$. For the Te-doped system, the DFT results are averaged over Case1 and Case2.
  }
  \label{elastic_modulus}
\setlength{\tabcolsep}{0.1em}
  \begin{tabular}{|c|c|c|c|c|c|c|c|c|}
  \hline
\multirow{2}{*}{Modulus} & \multicolumn{5}{c|}{$\mathrm{Mo_3Sb_7}$} &  \multicolumn{3}{c|}{$\mathrm{Mo_3Sb_{5.5}Te_{1.5}}$}  \\
 \cline{2-9}
&   DFT & Exp.~100K & Ref.~\cite{Lindsay} 100K & Exp.~300K & Ref.~\cite{Lindsay} 300K& DFT  & Exp.~100K & Exp.~300K \\
\hline
C11 (GPa)	&	204	&	173$\pm$28	&	161	&	172$\pm$28	&	178	&	195	&	183$\pm$22	&	191$\pm$27	\\
C12 (GPa)	&	98	&	68$\pm$11	&	90	&	83$\pm$13	&	87	&	103	&	60$\pm$7	&	70$\pm$10	\\
C44 (GPa)	&	57	&	61$\pm$10	&	72	&	65$\pm$10	&	68	&	92	&	82$\pm$10	&	82$\pm$11	\\
$K_T$ (GPa)	&	134	&	103$\pm$16	&	114	&	112$\pm$18	&	117	&	133	&	100$\pm$12	&	110$\pm$15	\\
\hline
  \end{tabular}
\end{table*}

Using the phonon group velocities determined from DFT (validated against IXS), we proceed to estimate the phonon relaxation times and phonon mean free paths for the two compositions. In this analysis, we use the $\bm{q}$-dependent phonon group velocity $v_g(\bm{q})$ of acoustic modes obtained from DFT simulations, for both the parent $\mathrm{Mo_3Sb_7}$ and doped $\mathrm{Mo_3Sb_{5.5}Te_{1.5}}$. First, we estimate an average relaxation time $ \overline{\tau}$ and average mean-free-path $ \overline{\Lambda}$ on a $41\times41\times41$ ${\bm{q-}}$point mesh, using the expression for the lattice thermal conductivity in the relaxation time approximation, $\overline{\tau} = \frac{1}{3} \overline{C_vv_g^2(\bm{q}}) / \kappa_{\rm L}$ and $\overline{\Lambda} = \frac{1}{3} \overline{C_vv_g(\bm{q})} / \kappa_{\rm L}$, and $\kappa_{\rm L}$ from values reported by Candolfi \textit{et al.}~\cite{Candolfi_2009_1}. Results are reported in table~\ref{tau_and_lambda}. From these results, we can observe that the average phonon relaxation time and average mean-free-path both show a clear increase upon Te doping, especially at T=100K. The decrease in phonon linewidth $\Gamma_{\rm LW}$ at 300K, which is inversely proportional to the phonon relaxation time $\tau = 1/2\Gamma_{\rm LW}$, is also observed in our experimental IXS data, as can be seen in figure~\ref{linewidth}. Our measurements of phonon linewidths at T=100K for the Te-doped sample are limited by the experimental resolution, and doping effects cannot be quantified at this temperature. {An upper-bound estimate for the maximum phonon linewidth in the Te-doped sample at 100\,K is $0.9\pm0.2$\,meV along [0\,K\,0] for K = 0.2\,r.l.u.} Moreover, the 20\% to 45\% change in experimental $\Gamma_{\rm LW}$ or $\tau$ with Te doping along the measured high symmetry directions at 300K is in generally good agreement with our estimate of the change in average relaxation time in table~\ref{tau_and_lambda}, i.e., $\{\tau^{300\rm\,K}_{\rm doped}-\tau^{300\rm\,K}_{\rm undoped}\}/\tau^{300\rm\,K}_{\rm undoped} = 0.41$. From the data in tables~\ref{gv_comparison}, \ref{elastic_modulus}, and \ref{tau_and_lambda}, we can assess the respective contributions of the group velocities and lifetimes to the Te-induced increase in $\kappa_{\rm L}$. We find that at 100\,K, the larger phonon group velocities in the Te-doped material accounts for just $\sim$25\% of the increase in $\kappa_{\rm L}$, while the main effect is due to the increase in phonon relaxation times. However, at 300\,K the stiffening of phonon group velocities upon doping accounts for $\sim$70\% of the increase in $\kappa_{\rm L}$ (this figure assumes that the doping-induced stiffening is independent of temperature, which is well justified in view of Fig.~\ref{phonon_softening}-f,g,h). This can be understood as a larger effect of electron-phonon scattering on the phonon relaxation times at low temperature, leading to the large suppression in $\kappa_{\rm L}$ at 100\,K, as discussed more quantitatively below.

\begin{table*}
  \caption{Average phonon relaxation time ($\tau$) and average mean-free-path ($\Lambda$) calculated from expression of lattice thermal conductivity $\kappa_{\rm L} = \frac{1}{3}C_v v_g^2(\bm{q})\tau = \frac{1}{3}C_v v_g(\bm{q})\Lambda$, where group velocity at given $\bm{q}$, $v_g(\bm{q})$, and specific heat capacity at constant volume, $C_v$, have been obtained from DFT simulations for $\mathrm{Mo_3Sb_7}$ and $\mathrm{Mo_3Sb_{5.5}Te_{1.5}}$ at T = 300K and 100K. The lattice thermal conductivity ($\kappa_{\rm L}$) values are taken from Ref.~\cite{Candolfi_2009_1}}
  \label{tau_and_lambda}
    \setlength{\tabcolsep}{1em}
  \begin{tabular}{|c|c|c|c|c|c|c|}

  \hline
 & \multicolumn{2}{c|}{$\mathrm{Mo_3Sb_7}$} &  \multicolumn{4}{c|}{$\mathrm{Mo_3Sb_{5.5}Te_{1.5}}$}  \\
 \cline{2-7}
 & & & \multicolumn{2}{c|}{Case1} & \multicolumn{2}{c|}{Case2}\\
 \cline{4-7}
T in Kelvin	&	100	&	300	&	100	&	300	&	100	&	300	\\
Average relaxation time ($\tau$) in ps	&	10.3	&	18.3	&	40.2	&	25.3	&	41.7	&	26.2	\\
Average mean-free-path ($\Lambda$) in $\mathrm{\mathring{A}}$	&	228	&	401	&	1160	&	727	&	1190	&	746	\\
\hline
  \end{tabular}

\end{table*}

\begin{figure*}
\begin{center}
\includegraphics[trim=0cm 0.0cm 0cm 0.0cm, clip=true, width=0.9\textwidth]{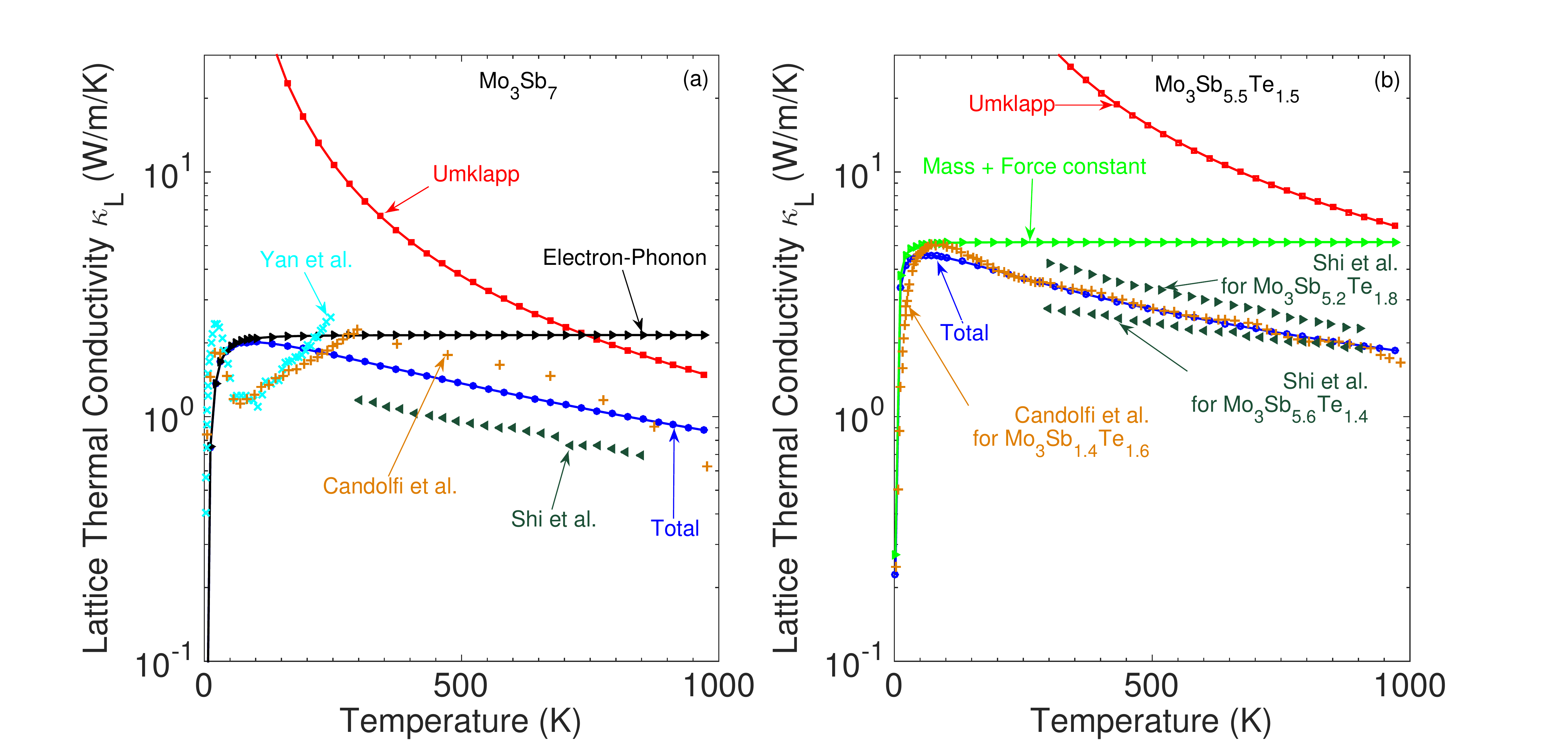}
\end{center}
\caption{\label{kappa} Model of the lattice thermal conductivity $\kappa_{\rm L}$ of a) $\mathrm{Mo_3Sb_7}$ and b) $\mathrm{Mo_3Sb_{5.5}Te_{1.5}}$, considering several phonon scattering processes, as described in the text. `$\times$' (cyan color), `$\lhd$' or `$\rhd$' (dark green), and `+' (brown color) are lattice thermal conductivity data from Yan \textit{et al.}~\cite{Yan_2013}, Shi \textit{et al.}~\cite{Shi_2011}, and Candolfi \textit{et al.}~\cite{Candolfi_2009_1,Candolfi_2010}, respectively. In b) `+' (brown color), `$\lhd$' (dark green), and `$\rhd$' (dark green) data points are for $\mathrm{Mo_3Sb_{5.4}Te_{1.6}}$, $\mathrm{Mo_3Sb_{5.6}Te_{1.4}}$, and $\mathrm{Mo_3Sb_{5.2}Te_{1.8}}$ composition.}
\end{figure*}

Several scattering mechanisms contribute to the overall phonon scattering rates, $\tau^{-1}$, which, following Matthiessen's rule, can be summed if assumed to be independent processes. In the present analysis, we consider boundary scattering ($\tau_B^{-1}$), point-defect scattering ($\tau_{PD}^{-1}$), umklapp scattering from phonon anharmonicity ($\tau_U^{-1}$), and electron-phonon scattering ($\tau_{ep}^{-1}$):
\begin{align}\label{tau1}
\tau^{-1} &= \tau_B^{-1} + \tau_{PD}^{-1} + \tau_U^{-1} + \tau_{ep}^{-1}. 
\end{align}
\begin{align}\label{tau2}  
\tau^{-1} &= \frac{v_g}{L} + 3V\frac{\omega^4}{\pi v_g^3}S^2 \nonumber \\
&+ P\frac{\hbar\gamma^2}{M v_g^2\theta_D}\omega^2T^n\exp\left(-\frac{\theta_D}{mT}\right) + C_{ep}\frac{\omega^2}{v_g^2},
\end{align}
where $L$ is the grain size, $V$ is the volume per unit cell, $\theta_D$ is the Debye temperature, $\omega$ is the phonon frequency, $\gamma$ is an average Gr\"{u}neisen parameter, and $P$, $n$, and $m$ are umklapp scattering parameters of order one. Scattering by point-defects, $S$, is estimated as a combination of three terms -- perturbations in mass, force-constant, and nearest-neighbor distance. The expression of $S^2$ and $C_{ep}$ are given by:\cite{Klemens,Tritt}
\begin{align}
S^2 &= S_1^2 + (S_2+S_3)^2, \nonumber \\
S_1 &= \frac{1}{\sqrt{12}}\left(\frac{\Delta M}{M}\right),\nonumber \\
S_2 &= \frac{1}{\sqrt{6}}\left(\frac{\Delta f}{f}\right),\nonumber \\
S_3 &= \sqrt{24}\left(\frac{\Delta R}{R}\right), \text{ and}\nonumber \\
C_{ep} &= \frac{4\rho m^*v_FL_e}{15d}.
\end{align}
Here, $\Delta M$, $\Delta f$, and $\Delta R$ are the perturbation in mass, force-constants, and nearest-neighbor distances, $\rho$ is the electron concentration, $m^*$ is the effective mass of electrons, $v_F$ is the Fermi velocity, $L_e$ is the mean free path of electrons, and $d$ is the mass density. The change in nearest neighbor distance (or local distortions) with Te doping is less than $0.1\%$, and is not considered in this analysis as it is much smaller than the perturbations in mass and force-constants. Debye temperatures $\theta_D = 282$\,K for $\mathrm{Mo_3Sb_7}$ and $\theta_D = 326$\,K for $\mathrm{Mo_3Sb_{5.4}Te_{1.6}}$ were estimated by Shi \textit{et al.} from their heat capacity measurements \cite{Shi_2011}. We have calculated the Debye temperature from the first moment of the measured phonon spectrum $g(E)$, as $\theta_{\rm D} = \frac{4\langle E\rangle}{3 k_{\rm B}}$, where $\langle E\rangle = \int g(E)E~dE$ and $k_{\rm B}$ is Boltzmann's constant. For $\mathrm{Mo_3Sb_7}$, we estimate Debye temperatures of $295\pm3$\,K and $290\pm3$\,K at 100\,K and 300\,K, respectively. Similarly, for $\mathrm{Mo_3Sb_{5.5}Te_{1.5}}$ we estimate $\theta_{\rm D}=312\pm4$\,K and $308\pm4$\,K at 100\,K and 300\,K, respectively. We used the reported value of grain size ($\sim$100\,$\mu m$) in thermal conductivity measurements \cite{Candolfi_2009_1,Candolfi_2010}, to estimate $L$ for boundary scattering. The parameters $P$, $n$, and $m$ have been kept constant at 1.0, 1.25, and 3.0, respectively, for both compositions. For $\mathrm{Mo_3Sb_7}$, we do not include any mass or force-constant disorder ($S=0$), while for $\mathrm{Mo_3Sb_{5.5}Te_{1.5}}$, the electron-phonon coupling constant $C_{ep}$ is set to zero (justified based on the much smaller carrier concentration). Based on mass and force-constant perturbations with Te doping (see table~\ref{BvK_comparison}), the terms $S_1$ and $S_2$ are  $3.70\times10^{-3}$ and $4.06\times10^{-2}$, respectively. Based on $\kappa_{\rm L}$ values reported by Candolfi \textit{et al.}~\cite{Candolfi_2009_1,Candolfi_2010}, the value of $C_{ep}$ for $\mathrm{Mo_3Sb_7}$ was estimated to be $1.00\times10^{-8}$ $\rm{m^2/sec}$. The $\bm q$- and branch-dependent phonon frequencies and group velocities in all of our lattice thermal conductivity calculations are determined from DFT simulations as described earlier.

The resulting lattice thermal conductivities $\kappa_{\rm L}(T)$ for the pure and doped systems are shown in figure~\ref{kappa}-a,b, respectively. Our results for the doped composition at all temperatures are in fair agreement with values of $\kappa_{\rm L}$ reported by Candolfi \textit{et al.}~\cite{Candolfi_2009_1,Candolfi_2010}. As can be observed, in $\mathrm{Mo_3Sb_{5.5}Te_{1.5}}$ the magnitude and temperature of the peak in $\kappa_{\rm L}$ is controlled by point-defect scattering processes at low temperatures, and at high temperatures Umklapp processes become large and determine the slope of lattice thermal conductivity. The dominance of Umklapp scattering at high temperature is consistent with the analysis of $\kappa_{\rm L}$ by Shi et al.~\cite{Shi_2011}. However, the prior study by Shi et al.~\cite{Shi_2011} underestimated the point-defect scattering by only considering the contribution from mass perturbation and ignoring the effect of force constant perturbation in $\mathrm{Mo_3Sb_{5.5}Te_{1.5}}$. However, according to our estimates, the force-constant perturbation ($S_2 = 4.06\times10^{-2}$) is approximately ten times greater than the mass perturbation ($S_1 = 3.70\times10^{-3}$), and cannot be ignored.

The very low lattice thermal conductivity in $\mathrm{Mo_3Sb_7}$ is a result of strong electron-phonon coupling in this compound, {combined with the large unit cell leading to many flat optical branches}. From figure~\ref{kappa}(a), we can observe that $\kappa_{\rm L}$ is greatly suppressed by scattering processes from electron-phonon interaction at low temperature, which still  contribute significantly at room temperature. As described earlier, the electron-phonon interaction affects both the phonon group velocities and the relaxation times. From our calculations, we find that in $\mathrm{Mo_3Sb_7}$ at 100K, the suppression in the phonon relaxation time compared to $\mathrm{Mo_3Sb_{5.5}Te_{1.5}}$ is dominated by the electron-phonon scattering, i.e., $\tau^{-1} \approx \tau_{ep}^{-1}$. Moreover, at 300K, electron-phonon scattering contributes about 85\% of the total phonon scattering rate compatible with large electron-phonon coupling in metals \cite{Grimvall1981}, while the increase in Umklapp scattering accounts for most of the rest. We note that we have ignored the interaction of phonons with Mo-Mo magnetic dimers and local distortions at low temperature, where magnetic properties may influence the thermal transport behavior. Consequently, our model does not capture the minimum in $\kappa_{\rm L}$ at $\sim55$\,K  reported by Candolfi \textit{et al.}~\cite{Candolfi_2009_1} and Yan \textit{et al.}~\cite{Yan_2013}. We attempted to include this extra source of phonon scattering with a resonance-scattering term \cite{Pohl_1962}, but the pronounced dip in experimental curves could not be reproduced. It is likely that the unusual $\kappa_{\rm L}$  of  $\mathrm{Mo_3Sb_7}$ near $\sim55$\,K is closely related to its structural instability induced by the competing magnetic interactions \cite{Yan_2015}, but a more complex phonon scattering model, beyond the scope of the present study, would be required to describe this effect. To summarize the present discussion, we emphasize that at temperatures ($200 \leqslant T \leqslant 750K$), the increase in $\kappa_{\rm L}$ with Te doping is primarily arising from the reduction in electron-phonon coupling, which significant increases the phonon group velocities and relaxation times.

\section{Conclusion}

We have investigated phonons in the candidate thermoelectric material $\mathrm{Mo_3Sb_{7-x}Te_x}$, using inelastic neutron and x-ray scattering, as well as DFT simulations. The phonon DOS exhibits a strong overall stiffening upon alloying with Te, which is well reproduced in our DFT simulations of alloy supercells. We also observed that the phonon DOS for both $\mathrm{Mo_3Sb_7}$ and $\mathrm{Mo_3Sb_{5.3}Te_{1.7}}$ shows a significant softening with increasing temperature, beyond the effect of thermal expansion in the quasi-harmonic approximation. Our first-principles phonon simulations for $\mathrm{Mo_3Sb_{7-x}Te_x}$ reveal that Te doping leads to strong local perturbations of the bonding environment and interatomic force-constants. In particular, the bonds Sb(12d)-Sb/Te(12d), Mo(12e)-Mo(12e), and Sb(16f)-Sb(16f) stiffen by as much as $\sim$108\%, 65\%, and 45\%, respectively, upon substitution of Te at the Sb(12d) site. As a result of this strong local perturbation, the phonon spectrum stiffens in a non-uniform manner. Furthermore, our phonon dispersion measurements with IXS, combined with DFT simulations, reveal that the suppression in electron-phonon coupling upon Te-doping largely accounts for the increase in lattice thermal conductivity. This suppressed electron-phonon coupling increases the phonon group velocities and phonon relaxation times, despite the introduction of significant impurity scattering from force-constant disorder. 

\section{Acknowledgements}
We would like to thank Dr.~John Tischler for providing software and guidance to fit the IXS spectra. Neutron and x-ray scattering measurements and analysis was supported by the U.S. Department of Energy, Office of Science, Basic Energy Sciences, Materials Sciences and Engineering Division, through the Office of Science Early Career Research Program (PI Delaire). Computer simulations and analysis were supported through CAMM, funded by the U.S. Department of Energy, Basic Energy Sciences, Materials Sciences and Engineering Division. Sample synthesis was supported by the U.S. Department of Energy, Office of Science, Basic Energy Sciences, Materials Sciences and Engineering Division. The use of Oak Ridge National Laboratory's Spallation Neutron Source was sponsored by the Scientific User Facilities Division, Office of Basic Energy Sciences, U.S. DOE. This research used resources of the Advanced Photon Source, a U.S. Department of Energy (DOE) Office of Science User Facility operated for the DOE Office of Science by Argonne National Laboratory under Contract No. DE-AC02-06CH11357. Theoretical calculations were performed using resources of the National Energy Research Scientific Computing Center, a DOE Office of Science User Facility supported by the Office of Science of the US Department of Energy under contract no.DE-AC02-05CH11231. 

\pagebreak



\end{document}